\documentclass[aps,pra,twocolumn,a4paper,superscriptaddress,amsmath,amssymb,floatfix]{revtex4-2}
\usepackage{graphicx,epstopdf}
\usepackage{enumerate}
\usepackage{braket}
\usepackage[colorlinks]{hyperref}
\hypersetup{%
  linkcolor=blue,
  citecolor=blue,
  filecolor=black,
  menucolor=black,
  urlcolor =black
}

\newcommand{\change}[1]{\textcolor{black}{#1}}

\newcommand{\Op}[1]{\boldsymbol{\mathbf{\hat{#1}}}}
\def\openone{\leavevmode\hbox{\small1\kern-3.3pt\normalsize1}}

\begin{document}

\author{J. Martin Berglund}
\affiliation{Theoretische Physik, 
Universit\"{a}t  Kassel, Heinrich-Plett-Stra{\ss}e 40,
34132 Kassel, Germany}
\affiliation{Instituto de física, 
Universidad Autónoma de San Luis Potosí, UASLP, Av. Parque Chapultepec 1570, Privadas del Pedregal, 78295 San Luis Potosí, S.L.P.}

\author{Michael Drewsen}
\affiliation{Department of Physics and Astronomy, 
Aarhus University, Ny Munkegade 120, DK-8000 Aarhus, Denmark}

\author{Christiane P. Koch}
\email{E-mail: christiane.koch@fu-berlin.de}
\affiliation{Theoretische Physik, 
Universit\"{a}t  Kassel, Heinrich-Plett-Stra{\ss}e 40,
34132 Kassel, Germany}
\affiliation{Freie Universit\"{a}t Berlin, 
Fachbereich Physik \& Dahlem Center for Complex Quantum Systems, Arnimallee 14, 14195 Berlin, Germany}

\title{Rotational state changes in collisions of diatomic molecular ions with
atomic ions}

\date{\today}

\begin{abstract}
We investigate rotational state changes in a single collision of diatomic
molecular ions, polar or apolar, with an atomic ion. 
\change{Rotational state changes may occur since the angular degree of freedom of the molecular ions interacts with the electric field due to the atomic ion.} Thanks to the very different time and energy scales
of translational and rotational motion, we may treat the collision
classically and describe \change{only the} rotations quantum mechanically. 
\change{We first investigate a number of example systems 
numerically and then} derive closed-form
\change{approximations} for the rotational excitation per collision, depending on the scattering energy and the molecular parameters. 
\change{These findings provide the basis for estimating the accumulated rotational excitation in sympathetic cooling of molecular ions by laser-cooled atomic ions [arXiv:2410.22458 ] which involves many single collisions.}
\end{abstract}

\maketitle

\section{Introduction}\label{sec:Intro}
Cold molecule science is a growing field of research with \change{applications}
 ranging from the test of fundamental physics, chemistry in the ultra-cold regime, to quantum information processing~\cite{BohnScience2017,OsterwalderBook2018,KremsBook18}.
Molecular ions, \change{either by themselves or } as hybrid systems together with neutral species \change{are proposed} for fundamental physics and quantum technological applications~\cite{TongPRL10, StaanumNatPhys10, HansenNature14}.  \change{For example, non-demolition quantum state detection experiments testify to the pristine level of control over molecular ions~\cite{Willitsch20}. A particular appeal of molecular ions is that} essentially any molecular ion with initial kinetic energy of $10\,$eV or lower can be trapped and sympathetically cooled through the Coulomb interaction with laser-cooled atomic ions~\cite{MolhavePRA00}. 
A variety of diatomic polar~\cite{MolhavePRA00,KoelemeijPRL07,StaanumPRL08,WillitschPRL08,HansenAngewandte12}
and apolar~\cite{BlythePRL05,TongPRL10} as well as larger molecular ions~\cite{OstendorfPRL06,HojbjerrePRA08} have been cooled this way to a few tens of millikelvin. 

\change{A typical scenario prepares the }
molecules in their rovibrational ground state~\footnote{We do not expect the situation for diatomic molecular ions prepared in rotationally excited states to be very different than the cases considered here. Changes in the magnetic sub-state populations can, however, occur depending on the orientation of the Coulomb-field, in analogy with previously discussed trap rf field induced transitions~\cite{HashemlooJCP15, HashemlooIJMPC16}.}, \change{for example by state-selective ionization. The molecular ions can be produced inside an ion trap or injected into the trap with the atomic ions where they then undergo collisions with the atomic ions.
If the collisions are elastic, the molecules transfer their kinetic energy to the atoms which are continuously laser-cooled, without any change to their internal state. However, the collisions may also be inelastic, changing the internal molecular state and reducing the quantum purity. This would impede use of the molecular ions in the envisioned applications in quantum information or tests of fundamental physics.}

\change{Naively one may expect no perturbation of the internal molecular state, since even at relatively high collisional energies of about $1\,$eV, corresponding to a temperature of about $10.000\,$K, the ion-ion distance never decreases to $1\,$nm. Indeed, close-encounter collisions, where the distance between the pair becomes comparable to or smaller than the size of the particles’ wave functions, are energetically suppressed by the ion-ion repulsion. However, the long-range nature of the atomic ion's Coulomb field may still result in internal state changes of the molecular ion. This is our concern here.}

\change{Starting from a first principles model, we separate translational and rotational motion, leveraging the huge difference in their energy, resp. time, scales~\cite{PatschJPCL22}. This allows us to describe the translational motion classically, as in the familiar textbook treatment of colliding point particles. The time-dependence of the atom-molecule distance then translates into a time-dependent Coulomb interaction that becomes a time-dependent external field (of near-Lorentzian shape) for the rotational motion of the molecule.  Solving the time-dependent rotational Schrödinger equation, we determine the probability of rotational state excitations in a single collision and examine its dependence on the collision energy and impact parameter for both polar and apolar diatomic molecular ions. We find the rotational excitation to sensitively depend on the molecular parameters, i.e., mass and dipole, respectively quadrupole moment and polarizability anisotropy for polar and apolar molecular ions.
Our results suggest that collisions can be used to infer the value of these molecular parameters from experimental measurements.}

\change{Moreover, in order to derive general scaling laws, we leverage perturbative and adiabatic approximations for apolar and polar molecular ions, respectively. This allows us to provide estimates for the rotational excitation in a single collisions of an arbitrary molecular ion with an atomic ion. In the companion paper~\cite{BerglundPRA2}, we  use these estimates to  calculate the overall rotational excitation that will be accumulated over a complete cooling cycle for two different experimental scenarios, a molecular ion co-trapped with a single atomic ion and a molecular ion immersed in a Coulomb crystal of atomic ions.
} 

\change{The paper is organized as follows. In Section~\ref{sec:Mod}, we introduce the theoretical model leveraging the separation of the energy scales of translational and rotational degrees of freedom. We also discuss the time-dependence of the electric field due to the atomic ion as felt by the molecular ion. In Sections~\ref{sec:Apol} and~\ref{sec:Pol}, we present our results for rotational state changes in collisions of apolar, respectively polar molecular ions with atomic ions. We conclude in Section~\ref{sec:Con} with an outlook.}

\section{Model}\label{sec:Mod}

\begin{figure}[tb]
 \centering
\includegraphics[width=0.79\linewidth]{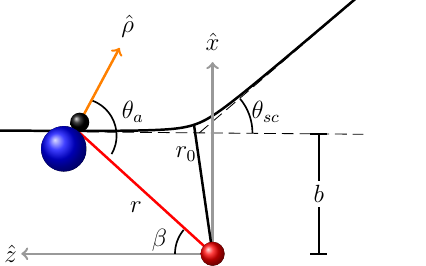}
 \caption{(Color online) Scattering geometry of a collision between a molecular ion and a laser-cooled atomic ion. Initially, the scattering pair are very far apart such that the vector $\vec{r}$ coincides with the 
 $\hat{z}$-axis \change{which thus also} defines the quantization axis ($\beta = 0$). $b$ is the impact parameter, $r_0$ the closest distance between the scattering pair and $\theta_{sc}$ the scattering angle. 
 $\theta_a$ is the angle between the molecular axis, $\hat{\rho}$, and $\vec{r}$. Since $\vec{r}$  changes its direction during scattering, the electric field \change{of the atomic ion, indicated in red} will acquire an $\hat{x}$-component for all but head-on collisions ($b=0$).}
 \label{fig:scheme}
\end{figure}
The energy scales of the  translational and rotational degrees of freedom are well separated: Initial scattering energies typically range from $0.1\,$eV to 10$\,$eV, whereas the rotational energy scale is only of the order of $10^{-4}\,$eV. \change{We therefore} can separate the two degrees of freedom. In particular, we \change{describe} the translational motion classically~\cite{HashemlooJCP15,HashemlooIJMPC16}, \change{as depicted schematically}  in Fig.~\ref{fig:scheme}. \change{Approximating the two ions as structureless point particles, the solution of the classical equation of motion is given in term's of Kepler's law.}
The rotational dynamics on the other hand, is described fully quantum mechanically,
\change{with the translational dynamics entering parametrically, as a time-dependent electric field. This is fully analogous to the parametric time-dependence of the electric dipole interaction between a Rydberg atom and a (neutral) polar molecule where the time-dependence arises from the classical trajectory of the inter-particle distance~\cite{PatschJPCL22}. In the present case, rotational state changes can occur since the dipole moment (in case of polar molecules), the quadrupole moment and the polarizability of the molecular ion are exposed to the electric field of the atomic ion. We assume, as is customary, that these moments are attached to the center of mass of the molecular ion. One may wonder whether in additional to rotational dynamics, the effectively time-dependent Coulomb interaction between the ions also affecs the molecular vibrations.}
Vibrational excitations may occur when the dipole moment changes with bond length, but the vibrational energy scale is typically about $100$ times larger than the rotational energy scale, and as we show in Appendix~\ref{sec:Avib}, the coupling is typically very weak. We therefore neglect the vibrational degree of freedom in our model, and only focus on the rotational degree of freedom, \change{treating the molecular ion as a rigid rotor}. 

\subsection{Classical description of the translational motion}\label{sec:Trans}
Considering the molecule as a point particle, its translational motion in the presence 
of the \change{atomic} ion reduces to a textbook scattering problem. 
At distances relevant for the scattering process, the \change{possible external trapping} potential can be neglected. 
The problem is then reduced to the classical scattering of a particle in a $1/r$-potential~\cite{Goldstein}, \change{cf.  Fig.~\ref{fig:scheme}}.
The scattering event is characterized by the scattering energy $E$ and the impact parameter. 
In order for \change{our assumption of energy scale separation to be} valid, we require the scattering energy to be  
low enough to ensure that the molecular and atomic ions' wave functions do not overlap. \change{This is equivalent to saying} that the smallest distance between the molecular ion and the coolant ion, or distance of closest approach $r_0$, needs to be significantly larger than the extension of the molecular ion. \change{Estimating the latter} by the equilibrium inter-atomic distance of the molecular ion, $r_e$, \change{this translates into} $r_0 \gg r_e$, where the distance of closest approach is
\begin{equation}\label{eq:r0} 
 r_0 = \frac{q_aq_m}{2E} + \sqrt{\left(\frac{q_aq_m}{2E}\right)^2 + b^2}.
\end{equation}
Here, $q_a$ and $q_m$ are the atomic and molecular charges. It follows from Eq.~\eqref{eq:r0} that $r_0 \ge q_aq_m/E$ at any given scattering energy $E$. \change{Taking the scattering energy $2\,$eV as an upper limit (as the typical highest initial collision energy in experiments on sympathetic translational cooling of molecular ions) and the example of MgH$^+$ with $r_e \approx 3.12\,a_0$ and $q_{at} = q_m = 1$ we find $r_0 \ge 13.5\,a_0$ with $a_0$ the Bohr radius. For lower energies, the assumption of non-overlapping wave functions is even better justified.} 

\change{Let us recall for completeness the textbook derivation of the trajectory $r(t)$ that describes the relative translational motion of atomic and molecular ion and determines the time-dependent field governing the molecule's rotational dynamics.}
With $M_{at}$ and $M_{mol}$ the atomic and molecular masses and $\mu = M_{at}M_{mol}/(M_{at}+M_{mol})$ the reduced mass, the scattering problem \change{of the two point particles} is described by the Hamiltonian
\begin{equation}\label{eq:keplerham}
  H = \frac{p_r^2}{2\mu} + V(r) + \frac{l^2}{2\mu r^2} = E,
\end{equation}
where $p_r$ is the radial momentum and $V(r)$ the 
Coulomb potential. The total energy $E$  
and the angular momentum $l$ are conserved quantities.
The angular momentum relates to the impact parameter $b$ by 
\begin{equation}\label{eq:lb}
 l = b p_0 = b\sqrt{2 \mu E}\,,
\end{equation}
where $p_0$ is the initial linear momentum of the molecule when $V(r)=0$
\change{(requiring very large separation due to the long-range nature of the Coulomb potential)}.
Solving for $p_r$ in  Eq.~\eqref{eq:keplerham} and using Eq.~\eqref{eq:lb} leads to \change{the familiar}
\begin{equation*}
  p_r = \sqrt{2\mu\left(E\left(1 - \frac{b^2}{r^2}\right) - V(r)\right)}\,,
\end{equation*}
\change{and with} $dr = v_rdt$ and $p_r=\mu v_r$, one arrives at
\begin{equation}\label{eq:timer}
  t(r) = \int_0^t dt' = \sqrt{\frac{\mu}{2}}\int_{r_0}^r\frac{dr'}
  {\sqrt{E\left(1 - \frac{b^2}{r'^2}\right) - V(r')}}\,,
\end{equation}
\change{from which, by inversion, $r(t)$ is obtained. Note that $t=0$ corresponds to $r=r_0$, i.e., the closest distance between the molecular and atomic ions.}

\change{When modeling the rotational dynamics in the presence of the electric  field due to the atomic ion below, it is important that,
for non-zero impact parameters, the field does not only change in magnitude.} During a collision, also its orientation with respect to the molecular ion changes. \change{We denote the angle between the field direction at time $t$ and the original field direction (the z-axis) by $\beta$,} see Fig.~\ref{fig:scheme}.
It is given by the classical trajectory
\change{~\cite{Goldstein}}, 
\begin{eqnarray*}
 \beta &=& \int_{r(t)}^{\infty}\frac{b\,ds}{s^2\sqrt{\left(1-\frac{V(s)}{E}\right)-\frac{b^2}{s^2}}}\;\;  \quad\mathrm{for}\; t<0\,,
 \end{eqnarray*}
\change{where we have replaced $r_0$ by $r(t)$ in the lower limit of the integral compared to the standard treatment~\cite{Goldstein},
and}
 \begin{eqnarray*}
  \beta &=& \int_{r_0}^{\infty}\frac{b\,ds}{s^2\sqrt{\left(1-\frac{V(s)}{E}\right)-\frac{b^2}{s^2}}} \\&&+ \int_{r_0}^{r(t)} \frac{b\,ds}{s^2\sqrt{\left(1-\frac{V(s)}{E}\right)-\frac{b^2}{s^2}}}
\quad\mathrm{for}\; t>0
\end{eqnarray*}
with $r(t) \in [r_0,\infty]$ and $r_0$ the minimal inter particle distance, cf. Fig.~\ref{fig:scheme}.  
\change{The role of $\beta$ is discussed in more detail in Section~\ref{sec:Qrot} below. The angle between the molecular axis and the electric field and the scattering angle are denoted by $\theta_a$ and $\theta_{sc}$ respectively, see Fig.~\ref{fig:scheme}. The scattering angle is given by~\cite{Goldstein}
\begin{equation*}
\theta_{sc} = \pi - \int_{r_0}^\infty\frac{b\,d\xi}{\xi\sqrt{\xi^2\left(1 - \frac{V(\xi)}{E}\right) -b^2 } }\,.
\end{equation*}
}

\subsection{The \change{magnitude of the} electric field seen by the molecular ion during a scattering event}\label{sec:Efield}

The \change{magnitude of the} electric field, seen by the molecular ion,  due to the Coulomb \change{interaction with the atomic ion} is given by 
\begin{equation}\label{eq:coulomb}
 \varepsilon(r(t)) = \frac{q_a}{r(t)^2}\,,
\end{equation}
taking its maximum value at the minimum distance, $r(t)=r_0$, \change{and Eq.~\eqref{eq:r0} implies
$\varepsilon_0 \to 1/b^2$. In other words, large impact parameter $b$ corresponds to low fields, as expected. In general, the electric field is a function of the time traversed, which in turn, according to Eq.~\eqref{eq:timer}, depends on the scattering energy $E$, the reduced mass $\mu$ and $b$, i.e., $\varepsilon = \varepsilon(t; E, \mu, b)$.} 

\begin{figure}[tb]
 \includegraphics[scale=0.35]{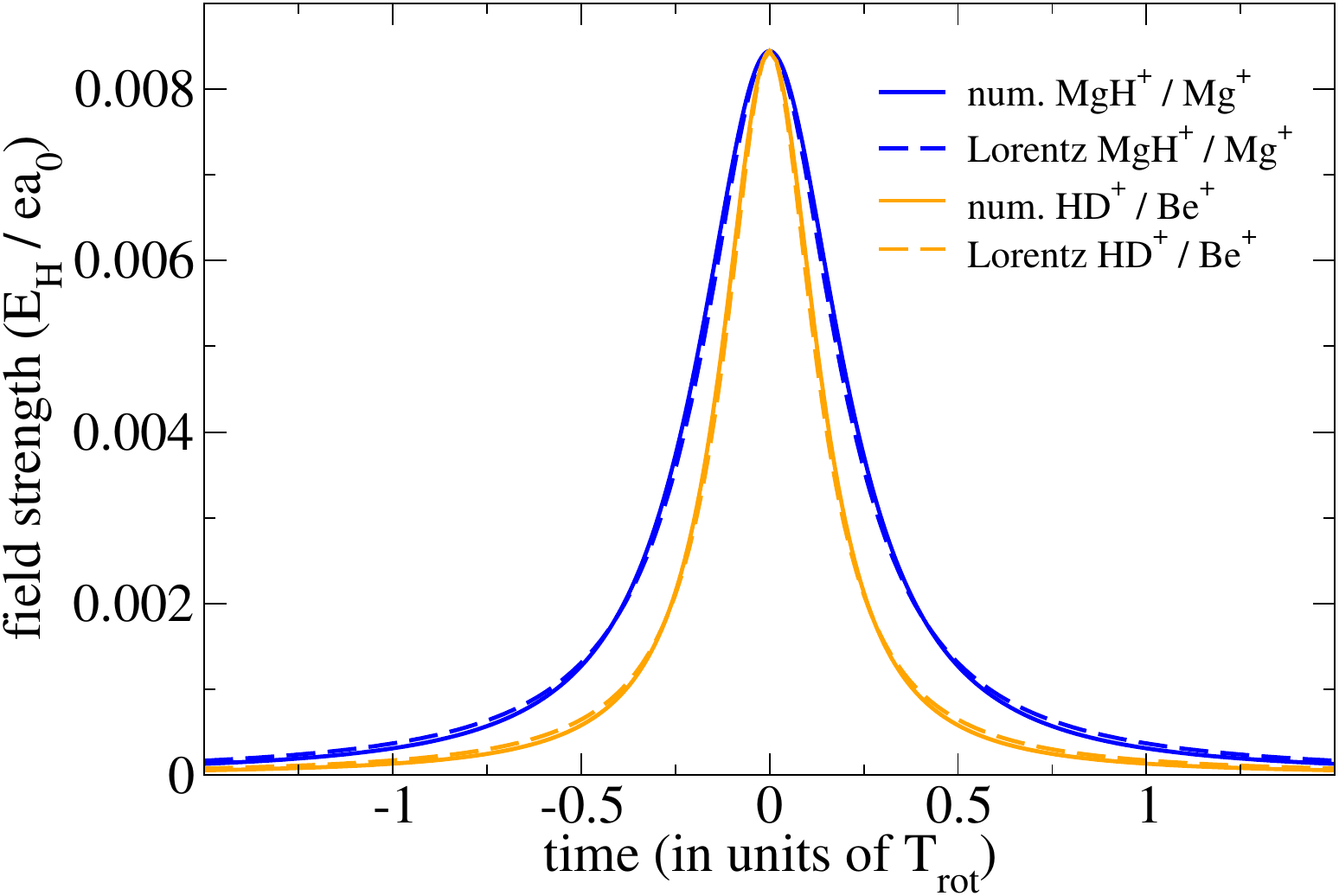}
 \caption{(Color online) The electric field \change{(in atomic units)} due to the \change{atomic} ion felt by the molecular ion for head-on ($b=0$) scattering at $E=2.5$~eV, \change{with the} Coulomb field transformed to a temporal field by Eq.~\eqref{eq:t_r1} (evaluated numerically),
 and compared to the Lorentz form, Eq.~\eqref{eq:Lorentzfield}.}
 \label{fig:Lorentzian}
\end{figure}

We first consider the case of head-on collisions ($b=0$).
\change{Since this yields the smallest $r_0$, cf. Eq.~\eqref{eq:r0}, one might expect} the largest excitation of the molecular rotational levels. 
In this case, the momentum is either parallel 
or anti parallel to the inter particle radius vector, cf. Fig.~\ref{fig:scheme}, and the angular momentum of
the scattering molecule is zero. 
With $V(r) = q_a/r$ the integral to be solved reduces to
\begin{equation}\label{eq:t_r1}
  t(r) = \int_0^t dt' = \sqrt{\frac{\mu r_0}{2\change{e^2}}}\int_{r_0}^r\frac{dr'}
  {\sqrt{1-r_0/r'}}.
\end{equation}
for single charged species ($q_aq_m=1\change{\cdot e^2}$) \change{as we consider here}.
The solution of the integral \eqref{eq:t_r1} has the form 
\begin{equation}\label{eq:int_sol}
t(r) = \sqrt{\frac{\mu r_0^3}{2\change{e^2}}} \left[ 
\frac{r}{r_0} \sqrt{1-r_0/r} + \frac{1}{2} \ln \left( 
\frac{\sqrt{1-r_0/r}+1}{|\sqrt{1-r_0/r}-1|}
\right)\right]\,, 
\end{equation}
where the prefactor $\tau/2 \equiv \sqrt{\mu r_0^3/(2\change{e^2})}$ has dimensions of time, and 
the first term is always the largest. For short distances, $r \sim r_0$, the second term can be approximated by a Taylor expansion around $r = r_0$, 
\begin{equation}\label{eq:tr_Taylor}
 t(r) = \frac{\tau}{2} \left(\frac{r}{r_0}\sqrt{1 - \frac{r_0}{r}} + \sqrt{1 - \frac{r_0}{r}}\right). 
\end{equation}
This can be rewritten, 
\begin{equation}\label{eq:tr3}
 \left(\frac{t(r)}{\frac{\tau}{2}}\right)^2 = \left(\frac{r}{r_0}\right)^2\left[\underbrace{\frac{5}{4} - \left(\frac{r_0}{r}-\frac{1}{2}\right)^2}_{a^2}\right] - 1\,,
\end{equation}
where the factor $a^2$ varies within the range $[1,1.25]$, and will hereafter be approximated by $a^2 = 1$.
Equation~\eqref{eq:tr3} can be rearranged into Lorentzian form,
\begin{equation}\label{eq:Lorentzfield}
 \varepsilon(t) = 
 \varepsilon_0\frac{\left(\frac{\tau}{2}\right)^2}{t^2+\left(\frac{\tau}{2}\right)^2}
\end{equation}
with the maximum electric field $\varepsilon_0 = \change{e}/r_0^2$ and the full width at half 
maximum (FWHM) given by $\tau = \sqrt{2\change{\mu r_0^3/e^2}} = \sqrt{2\mu\change{e^4}/E^3}$.
\change{The full, numerically calculated time dependence of the electric field during a head-on collision, taking the parameters for 
$^{24}$MgH$^+$ / $^{24}$Mg$^+$ and HD$^+$ / $^9$Be$^+$ , and a scattering energy of $2.5$ eV, is}  
plotted in Fig.~\ref{fig:Lorentzian}. \change{Fitting the curves to Lorentzians} leads to FWHM values given by 
\begin{equation}\label{eq:tau}
 \tau = 1.86\sqrt{\frac{\mu\change{e^4}}{E^3}}\,,
\end{equation}
where the prefactor differs from the analytical value $\sqrt{2}\approx 1.41$. \change{This is due to } the first order Taylor expansion going from Eq.~\eqref{eq:int_sol} to Eq.~\eqref{eq:tr_Taylor}, i.e., the true field is only approximately of Lorentzian form.

\subsection{Quantum description of the rotational motion}\label{sec:Qrot}
\begin{figure}[tbp]
 \centering
\includegraphics[width=0.8\linewidth]{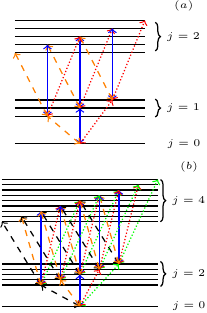}
 \caption{(Color online) Rotational excitations are due to the Coulomb field of the \change{atomic} ion coupling to the (a)
 dipole moment for polar molecular ions and (b) to the polarizability and quadrupole moment for apolar molecular ions. These different interactions  give rise to the different selection rules depicted in the two panels.  For head-on collisions only $\Delta m=0$ (blue arrows) are allowed.}
 \label{fig:trans}
\end{figure}

During the scattering process, the electric field of the atomic ion can affect the 
rotational dynamics of the molecular ion leading to excitation of its internal  
states. 
A standard electric multipole expansion of the field-ion interaction gives the three dominant terms as the monopole-, dipole- and quadruple interactions. The monopole term is responsible for the cooling of the translational degree of freedom, whereas the dipole- and quadrupole terms are responsible for the rotational dynamics.
\begin{table}[tb]
\begin{tabular}{c| c c c c c}
 Polar & $10^{-5}$B & $10^5$T$_{rot}$ & D & $Q_Z$ & $\mu$ \\
  \hline
  $^{24}$MgH$^+$ & 2.88 & 0.347 & 1.18 & 0.562 ($^*$)& 22473.21 \\
  HD$^+$ & 9.96  & 1.004 & 0.34 & $\approx 1.39$ ($^{**}$) & 4155.36 \\
\end{tabular}
\newline
\vspace*{0.25cm}
\newline
\begin{tabular}{c| c c c c c c}
 Apolar & $10^{-5}$B & $10^5$T$_{rot}$ & $\Delta\alpha$ & $\alpha_{\perp}$ & $Q_Z$ & $\mu$ \\
  \hline
  $^{14}$N$_2^+$ & 0.90 & 1.11 & 9.12 & 9.62 & 1.741 & 32463.57 \\
  H$_2^+$ & 12.69 & 0.079 & 3.72 & 1.71 & 1.39 & 3024.57 
\end{tabular}
\caption{Rotational constant $B$, dipole moment $D$, polarizability anisotropy $\Delta\alpha$, and quadrupole moment $Q_Z$ of a few molecular ions as well as reduced mass $\mu$ of molecular ion and coolant 
($^{24}$Mg$^+$ for MgH$^+$, $^9$Be$^+$ for HD$^+$, H$_2^+$, $^{48}$Ca$^+$ for N$_2^+$), all in atomic units. 
$^*$ very varying values between methods at NIST. 
$^{**}$ No values cited at NIST, we use the value given for H$_2^+$~\cite{NIST}.}
\label{tab:molpar}
\end{table}
Polar and apolar molecular ions couple differently to the external field. The polar molecular ions possess a permanent dipole moment $D$, whereas in apolar molecular ions $D$ vanishes due to symmetry.
The dipole interaction with the field $\varepsilon$ is simply $V_D= -D\varepsilon\cos{\theta_a}$, where $D$ is the dipole moment 
calculated at the equilibrium distance $r_e$ and $\theta_a$ is the angle between the electric field and the molecular axis, see Fig.~\ref{fig:scheme}. In the case of apolar molecular ions, with no permanent dipole moment, we consider the induced dipole via its polarizability. The interaction of the field with the induced dipole of a diatomic molecular ion is given by  
$V_{\alpha}=-\frac{\varepsilon^2}{4}\left(\Delta\alpha\cos^2{\theta_a} + \alpha_{\perp}\right)$ where $\Delta\alpha$ 
is the polarizability anisotropy. Finally the quadrupole interaction is $V_Q=\frac{Q_a\varepsilon^{3/2}}{4}\left(3\cos^2{\theta_a} + 1\right)$ where $Q_a$ 
is the permanent quadrupole moment of the molecule along the molecular axis.

The rotational kinetic energy is modeled within the rigid rotor 
approximation which leads to a kinetic energy term of $j(j+1)B$ where  $j$ is the rotational quantum number 
and $B \equiv \frac{1}{2\mu r_e^2}$ is the rotational constant, which sets a rotational timescale $T_{rot} = B^{-1}$. Together with Eq.~\eqref{eq:tau} we define a dimensionless time parameter,
\begin{equation}\label{eq:kappa}
 \kappa = B\tau = 1.86 \cdot B \sqrt{\frac{\mu\change{e^4}}{E^3}}
\end{equation}
such that $\kappa \ll 1$ ($\kappa \gg 1$) corresponds to a fast (slow) scattering regime. 

For polar molecules, the dipole interaction with the field dominates the rotational dynamics, and the Hamiltonian is given by 
\begin{eqnarray}\label{eq:ham1}
  \Op{H}_p &=& B\Op{J}^2 - D\varepsilon(t)\cos{\Op{\theta}_a} \\
  &=& B\Op{J}^2 - D\varepsilon(t)\left(\cos{\beta}\cos{\Op{\theta}} + 
  \sin{\beta}\sin{\Op{\theta}}\cos{\Op{\phi}}\right),\nonumber  
\end{eqnarray}
where $B$ is the rotational constant, $D$ the dipole moment, and 
$\theta_a$ the angle between the molecular axis and the electric field vector, cf. Fig.~\ref{fig:scheme}(a). \change{Note that $\theta_a$ is a function of the  angles $\beta$ (}between molecule and fixed scattering center in the CM frame\change{), 
, $\theta$ (between the molecular axis and the $\hat{z}$-axis, i.e. the initial electric field direction) and $\phi$ (the azimuthal angle around the molecular axis), all of which are time-dependent.} 

For apolar molecules, with no permanent dipole moment, the interaction is described by a Hamiltonian of the form
\begin{eqnarray}
\label{eq:ham2}
\Op{H}_{ap} &=& B\Op{J}^2 - \frac{\varepsilon^2(t)}{4}
\left(\Delta\alpha\cos^2{\Op{\theta}_a}  + \alpha_{\perp}\right)\nonumber\\
&&+ \frac{Q_{Z} \varepsilon^{3/2}(t)}{4}\left(3\cos^2{\Op{\theta}_a} + 1\right)\,,
\end{eqnarray}
where $\Op\theta_a$ can be substituted in terms of $\beta$, $\Op\theta$ and $\Op\phi$, similar to Eq.~\eqref{eq:ham1}. \change{As before, $\theta_a$ is a function of the time-dependent angles $\beta$, $\theta$ and $\phi$,} $\Delta\alpha$ is the polarizability anisotropy,  $\alpha_{\perp}$ 
the polarizability perpendicular to the molecular axis, and $Q_{Z}$ the quadrupole moment along the axis. 

The different field interactions lead to different selection rules for quantum transitions to excited rotational states. These are shown in Fig.~\ref{fig:trans} for (a) polar and (b) apolar molecular ions.
Notice that the nuclear wave function of apolar species is either even or odd with respect to rotations, and as a consequence allows only even or odd rotational states (c.f. ortho- and para hydrogen). In the case of odd states, panel (b) would be showing transitions between odd states. Since the spacing between the corresponding levels are significantly larger our model is expected to overestimate for odd nuclear spins.

The dynamics can be characterized in terms of the ratio of maximum interaction strength to rotational kinetic energy for the three types of coupling,
\begin{eqnarray}\label{eq:chi}
  \chi_D = \frac{D\varepsilon_0}{B}\,,\quad \label{eq:chiD}
  \chi_{\alpha} = \frac{\Delta\alpha\varepsilon_0^2}{4B},\quad
  \chi_Q = \frac{3Q_Z\varepsilon_0^{3/2}}{4B}    \,.
\end{eqnarray}

Consequently, if either ~$\chi_{D},\chi_\alpha,\chi_Q \ll 1$ we expect little excitation due to the given interaction, whereas a large value indicates that we can expect a large excitation.

For head-on collisions, when the interaction is strongest, $\frac{\chi_Q}{\chi_D} = \frac{3Q_Z}{4D}E$ is equal to 
$0.013E$ (in eV) for MgH$^+$ and $0.11E$ for
HD$^+$. In contrast, for apolar molecular ions  the quadrupole interaction dominates, 
for example, for N$_2^+$ at 2$\,$eV $\frac{\chi_Q}{\chi_{\alpha}} 
\approx 8$.
Additionally, the long-range behavior also favors the quadrupole interaction. 

Now that we have the Hamiltonians that generate the rotational dynamics in the presence of the time dependent electric field seen by the molecular ion in a scattering even, we will use them to solve the time dependent Schrödinger equation 
\begin{equation}\label{eq:TDSE}
i\hbar\partial_t\Ket{\psi(t)} = \Op{H}(t)\Ket{\psi(t)}
\end{equation}
to study the rotational excitation dynamics. \change{In order to solve Eq.~\eqref{eq:TDSE}, we represent Hamiltonian and wavefunction in the basis of spherical harmonics and discretize time
such that the Hamiltonian can be considered constant within one time step $\Delta t$. Due to the long range nature of the Coulomb potential, we use different time steps for the far-, mid- and close ranges. The formal solution of Eq.~\eqref{eq:TDSE}, 
$\Op{U}(\Delta t) = e^{-i\Op{H}(\tilde{t})\Delta t}$, where $\tilde{t}$ is the midpoint of the time interval, can be calculated with any desired precision by expanding the exponential in Chebychev polynomials~\cite{KosloffJCP84}.} Due to the different couplings to the external field for polar and apolar molecular ions we will consider them separately in the following discussion. 

\section{Scattering of apolar molecular ions}\label{sec:Apol}
We start our discussion with numerical integration of Eq.~\eqref{eq:TDSE} for apolar molecular ions. 

\subsection{Numerical solution of the time dependent Schrödinger equation}\label{sec:Anum}
\begin{figure}[tbp]
 \centering
 \includegraphics[width=0.95\linewidth]{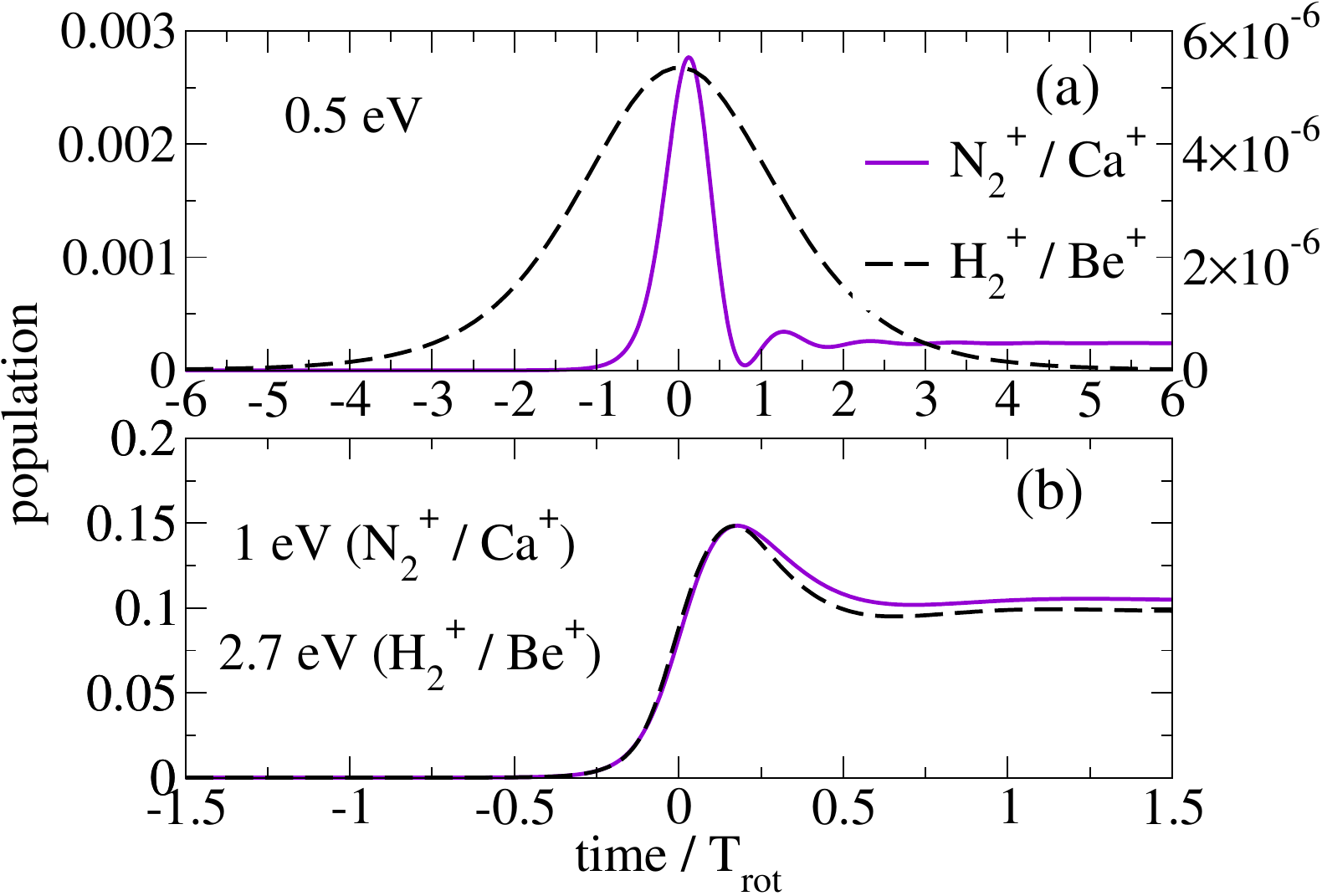}
\caption{(Color online) Population excitation for apolar molecular ions as a function of time of a head-on scattering event
  with energy $E$ indicated in the panels. (a) For low scattering energies the motion approaches the adiabatic limit, as is clearly seen for H$_2^+$ / Be$^+$-scattering. Here the left scale is for N$_2^+$ / Ca$^+$ and the right scale is for H$_2^+$ / Be$^+$. (b) For higher scattering energies adiabaticity is lost, and the excitation is to a large degree determined by the product $\chi_Q\kappa$. The dynamics is qualitatively the same for other apolar molecular species.}
\label{fig:dyn_pop2}
\end{figure}
We begin by studying the rotational dynamics at a particular scattering energy and head-on collisions, as presented in Fig.~\ref{fig:dyn_pop2}. .
In agreement with the different values of the parameters $\chi_{Q}/\chi_{\alpha}$, see Eq.~\eqref{eq:chi} and table \ref{tab:molpar}, the dynamics for N$_2^+$ and H$_2^+$ is very different at the same scattering energy, as seen in panel (a) of the figure. Notice that the timescale for H$_2^+$ is long compared to the rotational time, $T_{rot} = B^{-1}$, at this scattering energy, leading to suppression of the final excitation. Notice also the different y-axes for the different species in this panel. The dynamics for different apolar molecular ions can be almost identical, if the dynamics is studied at different energies, such that the product $\chi_Q\kappa$ is same for the two species, as is exemplified in panel (b). Notice the low to moderate intermediate and final population excitations associated with scattering of apolar molecular ions. The scattering dynamics is qualitatively the same for different apolar molecular ions. The scattering is also qualitatively the same for non-head-on scattering, with small excitations to $m\neq 0$-sub-levels, and the maximum excitation transfer takes place at head-on collisions.

\subsection{Perturbation theory treatment}\label{sec:Pert}
The small to moderate excitation dynamics associated with apolar molecular ions during scattering, suggest that first order perturbation theory (PT) can 
be applied to estimate their population excitation.
In this case, the final-time amplitude of the lowest excited rotational state $\Ket{2,0}$ 
after a single collision at energy $E$ and impact parameter $b$, is given by
\begin{equation}\label{eq:c_PT}
 c_{2,0}^{(1)} (E,b) = -i\int_{-\infty}^{\infty}\Braket{2,0|\Op{H}_{int}|0,0}e^{i6Bt}dt.
\end{equation}
The contribution to the population excitation due to the polarizability interaction in Hamiltonian~\eqref{eq:ham2} of Eq.~\eqref{eq:c_PT} yields   
\begin{widetext}
  \begin{equation}\label{eq:c_np0}
   c_{\Delta\alpha} (E,b) =-i\chi_{\alpha}(E,b)B\left(\frac{\tau(E)}{2}\right)^4\int_{-\infty}^{\infty}
   \frac{e^{i6Bt}}{\left(t^2+\left(\frac{\tau(E)}{2}\right)^2\right)^2}dt\Braket{2,0|\cos^2{\theta}|0,0}.
  \end{equation}
\end{widetext}
The parameter $\chi_Q(E,b)$, defined in Eq.~\eqref{eq:chi}, is a function of both the scattering energy 
and impact parameter through the maximum electric field strength $\varepsilon_0(E,b)$. Note that within our model it is the only 
quantity determining the population excitation that depends on the impact parameter $b$.
The integral Eq.~\eqref{eq:c_np0} is easily evaluated using Cauchy's integral formula for derivatives, with the result 
\begin{equation}\label{eq:c_np02}
 c_{\Delta\alpha}(\chi_{\alpha}, \kappa) = -i\frac{\pi}{4}\chi_{\alpha}\kappa(1+3\kappa)e^{-3\kappa}\Braket{2,0|\cos^2{\theta}|0,0}.
\end{equation}
Notice that the expansion coefficient depends only on $\kappa(E)$ and not on $\tau(E)$ and $B$ separately.

The contribution to the excitation due to the quadrupole interaction term of Hamiltonian~\eqref{eq:ham2} in Eq.~\eqref{eq:c_PT} yields 
\begin{widetext}
  \begin{equation}\label{eq:c_np1}
   c_Q (E,b) =i\chi_Q(E,b)B\left(\frac{\tau(E)}{2}\right)^3\int_{-\infty}^{\infty}
   \frac{e^{i6Bt}}{\left(t^2+\left(\frac{\tau(E)}{2}\right)^2\right)^{3/2}}dt\Braket{2,0|\cos^2{\theta}|0,0}.
  \end{equation}
\end{widetext}
Here the integral is not of the form of a Cauchy integral formula for derivatives. However, by the variable transformation
$t = \frac{\tau}{2}\tan{u}$, the integral over time in Eq.~\eqref{eq:c_np1} can be written as
\begin{equation}
\left(\frac{\tau}{2}\right)^2\int_{-\infty}^{\infty}\frac{e^{i6Bt}dt}{\left(t^2+\left(\frac{\tau}{2}\right)^2\right)^{3/2}}
=\int_{-\frac{\pi}{2}}^{\frac{\pi}{2}}\cos{u}e^{i3\kappa\tan{u}}du \equiv f(\kappa). 
\end{equation}
In this form we see explicitly that the value of the integral only depends on $\kappa(E)$ just as with the polarizability term, and it evaluates to
\begin{equation}\label{eq:fKappa}
 f(\kappa) \approx 2\sqrt{1+6\kappa}e^{-3\kappa},
\end{equation}
with the details presented in Appendix~\ref{sec:Aint}. Using the result of Eq.~\eqref{eq:fKappa} in Eq.~\eqref{eq:c_np1} we obtain the amplitude for excitation due to the quadrupole excitation as
\begin{equation}\label{eq:excQuad}
 c_{Q}(\chi_Q, \kappa) \approx i \chi_{\change{Q}} \kappa \sqrt{1+6\kappa} e^{-3\kappa}\Braket{2,0|\cos^2{\theta}|2,0}.
\end{equation}
The form of $f(\kappa)$ arises by assuming the ground state as initial state, and due to the selection rule $\Delta j = 2$ we have $\Delta E = 6B$. In general, with $\Delta j = 2$ we have $\Delta E_j = (4j+6)B$ for an initial state with rotational quantum number $j$. Consequently, in general $f(\kappa) \approx 2\sqrt{1 + (4j+6)\kappa}e^{(2j+3)\kappa}$. This treatment is of relevance e.g. when the even rotational states are forbidden by symmetry, in which case the ground states correspond to $j=1, \quad m = 0, \pm 1$ and consequently $\Delta E = 10B$. In this case we also have to consider the transition matrix elements $\Braket{3,m|\cos^2{\theta}|1,m}$, where $ m = 0, \pm 1$.

For both interactions the excitation depends on two parameters, related to molecular properties, namely $\chi_{\alpha/Q}$ and $\kappa$, and the scattering energy and the impact parameter. The relative contribution between the two interactions is given by the ratio 
\begin{equation}
 \frac{|c_{\Delta\alpha}|^2}{|c_Q|^2} = \left(\frac{\pi}{4}\right)^2 
 \left(\frac{\chi_{\Delta\alpha}}{\chi_Q}\right)^2 \frac{(1+3\kappa)^2}{1+6\kappa}
\end{equation}
At scattering energies relevant to us the quadrupole interaction is the dominant interaction. Scattering at even higher energies would eventually lead to dominating polarizability interaction, since 
$\frac{\chi_{\alpha}}{\chi_Q} \propto E^2$. At high energies (corresponding to low $\kappa$) we also have $\frac{(1+3\kappa)^2}{1+6\kappa} \to 1$. In the low $\kappa$-limit Eq.~\eqref{eq:excQuad} gives a population excitation proportional to $\left(\chi_Q\kappa\right)^2$ and adiabatic dynamics in the high $\kappa$-limit via the exponential factor, in agreement with Fig.~\ref{fig:dyn_pop2}.

\begin{figure}[tbp]
 \includegraphics[scale=0.35]{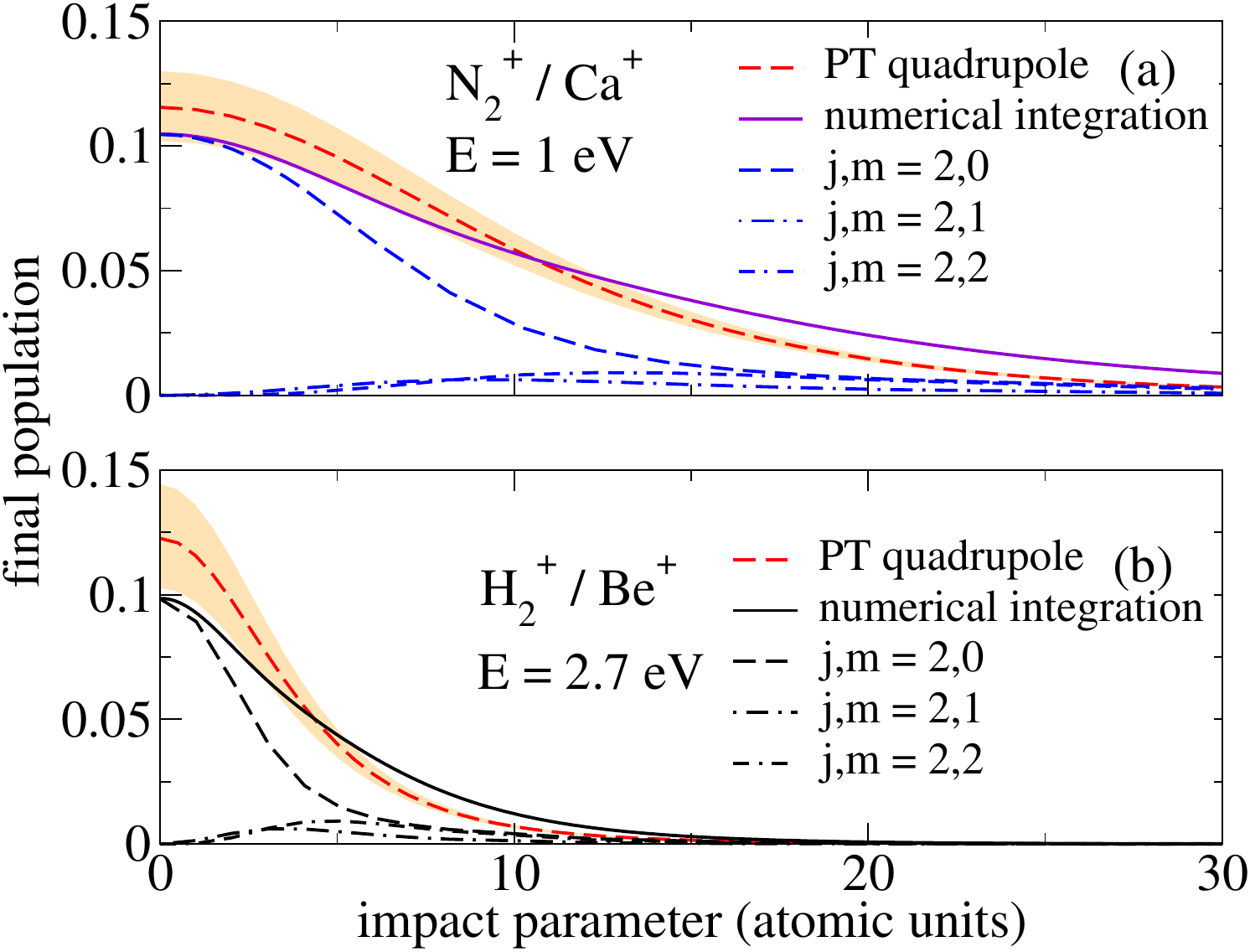}
 \caption{(Color online) Comparison between \change{final population excitation} as obtained from numerical integration of the Schr\"odinger equation generated by the Hamiltonian Eq.~\eqref{eq:ham2} and as obtained from PT taking only the quadrupole interaction into account (absolute square of Eq.~\eqref{eq:excQuad}). 
 The shaded region shows the maximum deviation due to the polarizability interaction, where the excitation takes values in $|c_Q|^2 + |c_{\Delta\alpha}|^2 \pm 2|c_Q||c_{\Delta\alpha}|$, with $|c_{\Delta\alpha}|$ the absolute square of Eq.~\eqref{eq:c_np02}. The solid lines show the full excitation given by numerical simulations, and the dashed and dotted lines show the final population on individual $m$-states for $j=2$.}
 \label{fig:chiPop}
\end{figure}
The actual population excitations can be faithfully obtained from perturbation theory including only the quadrupole interaction, c.f.~Fig.~\ref{fig:chiPop}.
The population excitation is shown as function of the impact parameter $b$ for the analytical expression of the quadrupole interaction 
(absolute square of Eq.~\eqref{eq:excQuad}
\change{which scales as $1/b^{6}$ for large $b$%
}) compared to the numerical simulation of the Schr\"odinger equation~\eqref{eq:ham2}. The analytical results are given
with margins (shaded region in the figure) obtained from including the polarizability interaction, $|c_Q|^2 + |c_{\Delta\alpha}|^2 \pm 2|c_Q||c_{\Delta\alpha}|$. 
The agreement between analytical and numerical results is good, in particular for low $b$. 
Notice that the margins increase for small $b$, as is expected since a larger field strength 
favors the polarizability interaction, which is reflected in the ratio $\frac{|c_{\Delta\alpha}|^2}{|c_Q|^2}$, which increases with increasing energy. Also note the larger shaded area for H$_2^+$, reflecting the relatively large polarizability interaction to be expected from the molecular parameters, see Table~\ref{tab:molpar}.
The more rapid decrease in excitation from the analytical- as compared to the full numerical calculations is likely due to the neglect of the transitions to 
non-zero $m$-states as suggested by Fig.~\ref{fig:chiPop}.
The maximum population transfer for apolar molecules does occur at $b=0$ in accordance with its maximum $\chi_Q$-value, Fig.~\ref{fig:chiPop}.  
Notice that most of the excited population occurs for the $m$-conserving part of the interaction resulting in the mayor excited 
population in the $m=0$ sub-level. The slight underestimation of the simulated excitation by our model can be attributed to the neglect of the excitation of the non-zero $m$-states and the approximation that $\tau$ is independent of $b$.

\section{Scattering of polar molecular ions}\label{sec:Pol}
We now turn our attention to solving Eq.~\eqref{eq:TDSE} for polar molecular ions.

\subsection{Numerical solution of the time dependent Schrödinger equation}\label{sec:Pnum}
\begin{figure}[tbp]
 \centering
 \includegraphics[scale=0.325]{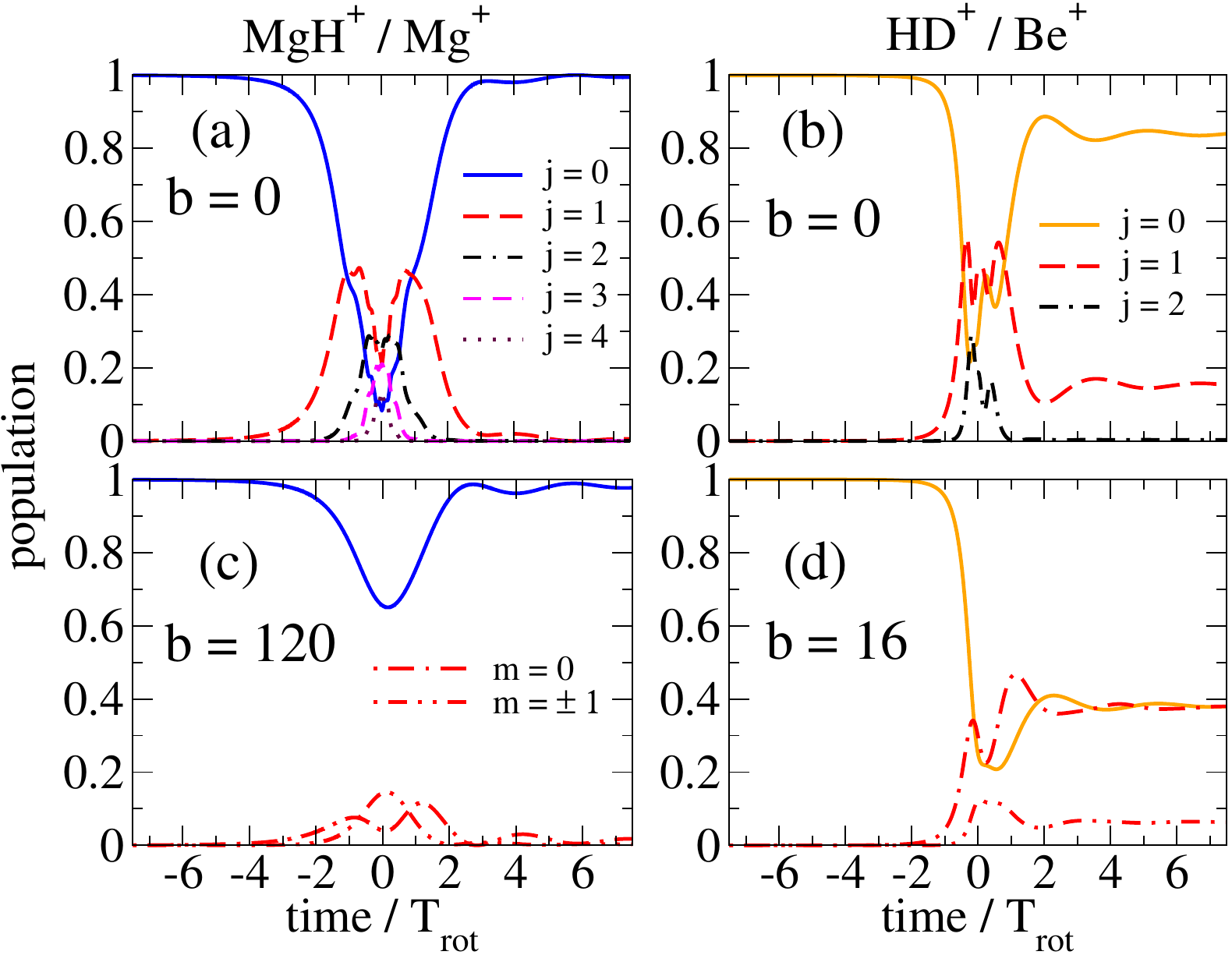}
 \caption{(Color online) Population dynamics for scattering energy $E = 2$~eV, for MgH$^+$ / Mg$^+$ (a,c) and  HD$^+$ / Be$^+$ (b,d). (a,b) show the dynamics at $b=0$, whereas (c,d) for the impact parameter that leads to the most population excitation.
  The collision \change{dynamics for head-on collisions is close to the adiabatic limit (a,b), wheras for non head-on collisions adiabaticity is gradually lost (c,d).} The dynamics is qualitatively the same for other polar molecular species.}
 \label{fig:dyn_pop_polar}
\end{figure}

We first consider the population excitation dependence in a scattering event at a fixed scattering energy.
Contrary to what would be expected for the largest $\chi_D$-value ($\chi_D \approx 220,18$ for MgH$^+$ and HD$^+$ respectively), 
numerical integration of the Schr\"odinger equation generated by the Hamiltonian Eq.~\eqref{eq:ham1} show, most strikingly, a low final population excitation, c.f. Fig.~\ref{fig:dyn_pop_polar}. In particular, one would expect that the excitation would be significantly more pronounced for MgH$^+$, but what actually occurs is a larger excitation for HD$^+$ ions, contrary to what the relative $\chi_D$-values of both ions suggests. However, although the final rotational state excitation is very small, 
the effect of the field on the rotational state dynamics is significant during the collision due to the large $\chi_D$-values associated with polar molecular ions. In particular, for MgH$^+$ at head-on collision, the intermediate excitation leads to population of several rotational levels and the ground state is temporarily all but depleted, but most of the population tends to return to the ground state after the collision event, (a). Notice the lower intermediate population transfer for HD$^+$ (b) as compared to MgH$^+$, but its larger final population transfer. This trend also holds for nonzero values of $b$, as shown for $b \approx 120, 16$ where the maximum final excitation 
occurs for MgH$^+$/HD$^+$ respectively (c,d). 

\begin{figure}[tbp]
\includegraphics[scale=0.28]{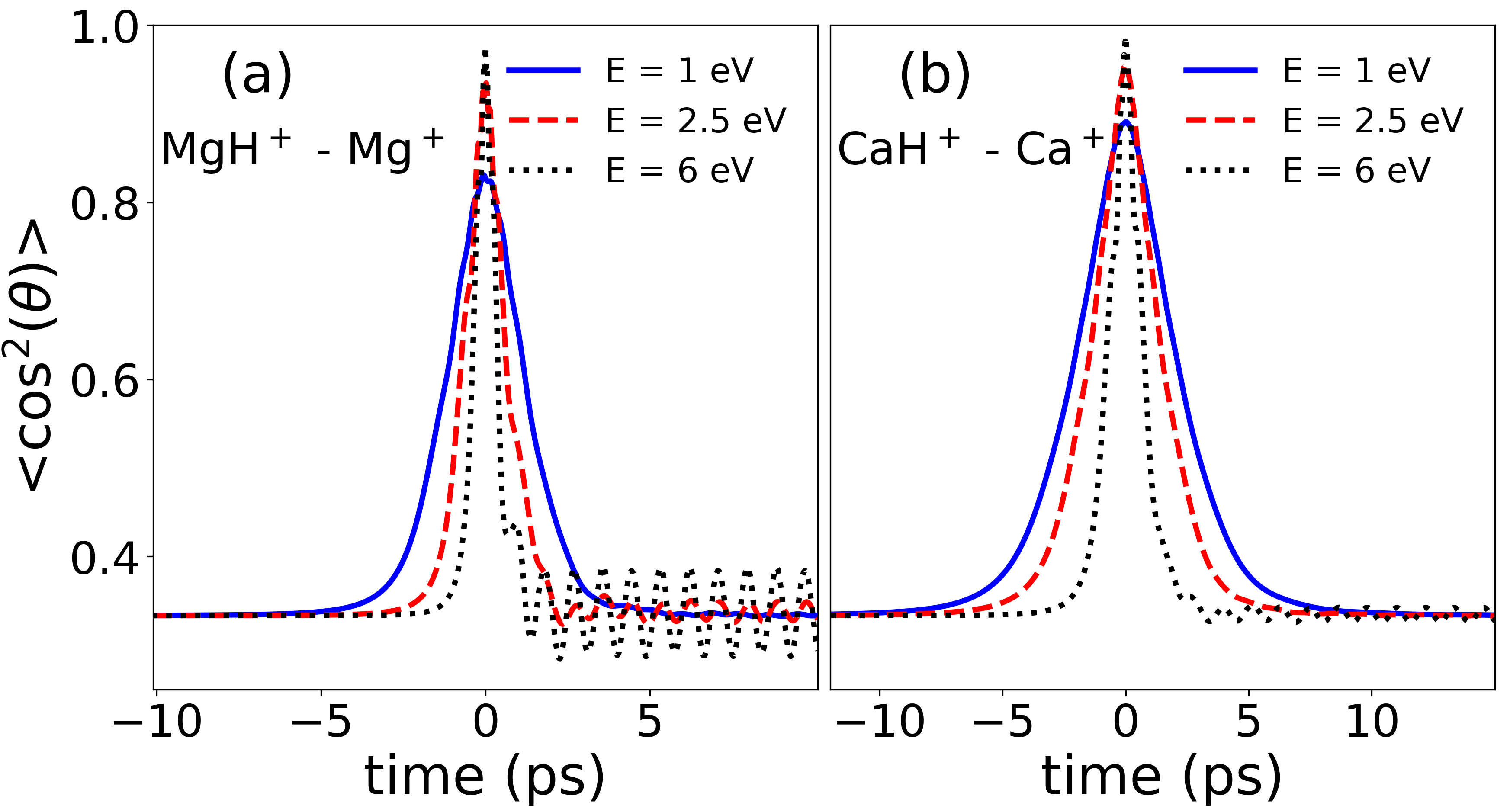}
\caption{The alignment factor $\Braket{\cos^2{\theta}}$ as function of time at head-on collisions at different scattering energies, $E$. The different panels show different scattering pairs.}
 \label{fig:polarmatrix}
\end{figure}

Therefore, no correlation between the value of $\chi_D$ and the final population excitation can be established, disregarding the impact parameter.  
The strong intermediate excitation, due to the high $\chi_D$-value, does, however cause a temporary alignment of the molecular ions, see Fig.~\ref{fig:polarmatrix}. The alignment, associated with the gradual on/offset of the Lorentzian, acts to prepare the molecular ion for the strong interaction near the center of the Lorentzian. Notice that the maximum alignment factor increases with increasing energy, whereas the scattering time decreases, since $\tau$ decreases with the scattering energy. The final population excitation does increase with increasing scattering energy. 
It is also clear that the maximum degree of orientation is not a relevant predictor for the final population excitation, where panels (a) and (b) indicate similar degree of maximal orientation at any given scattering energy (as a consequence of the high $\chi_D$-values associated with the scattering in all cases) and slightly larger in panel (b), whereas the largest final population transfer is seen in panel (a). The lower value of $\tau$ of the electric field associated with higher scattering energy correlates better to increasing final population transfer, suggesting that the timescale plays a significant role for the level of final excitation.

\begin{figure}[tbp]
\centering
\includegraphics[scale=1.0]{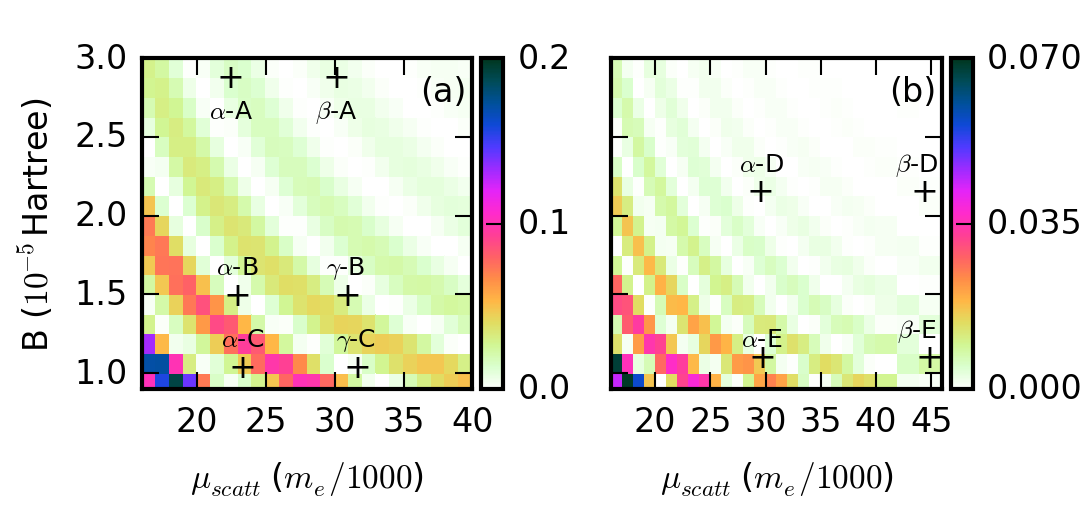}
\caption{Population excitation as function of the rotational constant, $B$, and the reduced mass, $\mu$, for two dipole couplings, $D=1.18$ (i.e. the dipole moment of MgH$^+$) in panel (a) and $D=2.34$ (i.e. the dipole moment of CaH$^+$) in panel (b). Scattering energy is $E=2.5$ eV. The coolants are denoted with Greek letters, $\alpha =$ Mg$^+$, $\beta =$ Ca$^+$, $\gamma = $ Ba$^+$. The molecular ions are designated Latin letters, $A = $MgH$^+$, $B = $MgD$^+$, $C = $MgT$^+$, $D = $CaH$^+$, $E =  $CaD$^+$.}
\label{fig:poplandscape}
\end{figure}

Even though the final excitation does not correlate to the $\chi_D$-value, numerical simulations suggest another correlation between molecular and scattering parameters, as the plots in Fig.~\ref{fig:poplandscape} shows. Here the population excitation is shown as a function of the rotational constant and reduced scattering mass at scattering energy $E=2.5$ eV and head-on collisions. In the left panel a lower dipole moment is used ($D=1.18$, corresponding to MgH$^+$), and a larger dipole moment in the right panel ($D=2.34$, corresponding to CaH$^+$). We first notice the difference in the amount of excitations in the two panels, where, surprisingly, least excitation occurs for the higher dipole moment. The excitation is also suppressed for large values of both the rotational constant and the reduced scattering mass, as is seen in both panels of the figure. There is also an oscillatory pattering in the final excitation, with increasing frequency for higher dipole moment, suggesting that a constant product $B\mu$ should lead to constant excitation. 

The combined results of the numerical simulations suggest a near adiabatic dynamics, particularly pronounced within the high field limit. We will therefore continue to compute an estimation for the final excitation in the frame of an adiabatic approximation.
\subsection{Treatment in the adiabatic picture}\label{sec:Adia}
In the adiabatic picture, the Hamiltonian changes slowly with respect to the energy scale set by the instantaneous eigenvalues of the Hamiltonian. Therefore, consider the instantaneous eigenstates $\Ket{\psi_\iota(t)}$ of a time-dependent Hamiltonian $\Op{H}(t)$ of the form \eqref{eq:ham1}, 
\begin{equation}\label{eq:Ad1}
 \Op{H}(t)\Ket{\psi_{\iota}(t)} = E_{\iota}(t)\Ket{\psi_{\iota}(t)},
\end{equation}
with eigenenergies $E_\iota(t)$. We will use the labelling $\iota,\iota'$ to denote the states of the adiabatic basis to distinguish them from the 
field free basis with labels $j,j'$.
Any state $\Ket{\Psi(t)}$ can be expanded into the time-dependent eigenstates, $\Ket{\Psi(t)} = \sum_\iota c_{\iota}(t)e^{i\Theta_{\iota}(t)}\Ket{\psi_{\iota}(t)}$, where $\Theta_{\iota} = -\int^tE_{\iota}(t')dt'$. Inserting the ansatz into the time-dependent Schr\"odinger equation, Eq.~\eqref{eq:TDSE}, leads to the expression for the expansion coefficient
\begin{widetext}
 \begin{equation}\label{eq:eom_ci}
  \dot{c}_{\iota '}(t) = -c_{\iota '}\Braket{\psi_{\iota '}(t)|\dot{\psi}_{\iota '}(t)} - \sum_{{\iota '} \neq {\iota}} c_{\iota}(t)e^{i\Delta\Theta_{\iota \iota '}(t)}\frac{\Braket{\psi_{\iota '}(t)|\partial_t\Op{H}(t)|\psi_{\iota}(t)}}{E_{\iota}(t)-E_{\iota '}(t)}
 \end{equation}
\end{widetext}
for an excited state (see the Appendix for details).

Since, by assumption we have near adiabatic dynamics    
$|c_0(t)| \sim 1$ and $|c_{\iota'}(t)| \ll 1, \iota' \neq 0$. Then integrating Eq.~\eqref{eq:eom_ci} term by term,
\begin{equation}\label{eq:cadia}
  c_{\iota'}(t) \approx \int_{-\infty}^t e^{i\Delta\Theta_{0\iota'}(t^\prime)}\frac{\Braket{\psi_{\iota'}(t^\prime)|\partial_t\Op{H}(t^\prime)|\psi_0(t^\prime)}}{E_{\iota'}(t^\prime)-E_0(t^\prime)}dt^\prime.
\end{equation}
In order to evaluate Eq.~\eqref{eq:cadia} we need to make further approximations. We therefore need to find some limits in which the evaluation of Eq.~\eqref{eq:cadia} is feasible. 

\begin{figure}[tbp]
 \includegraphics[scale=0.35]{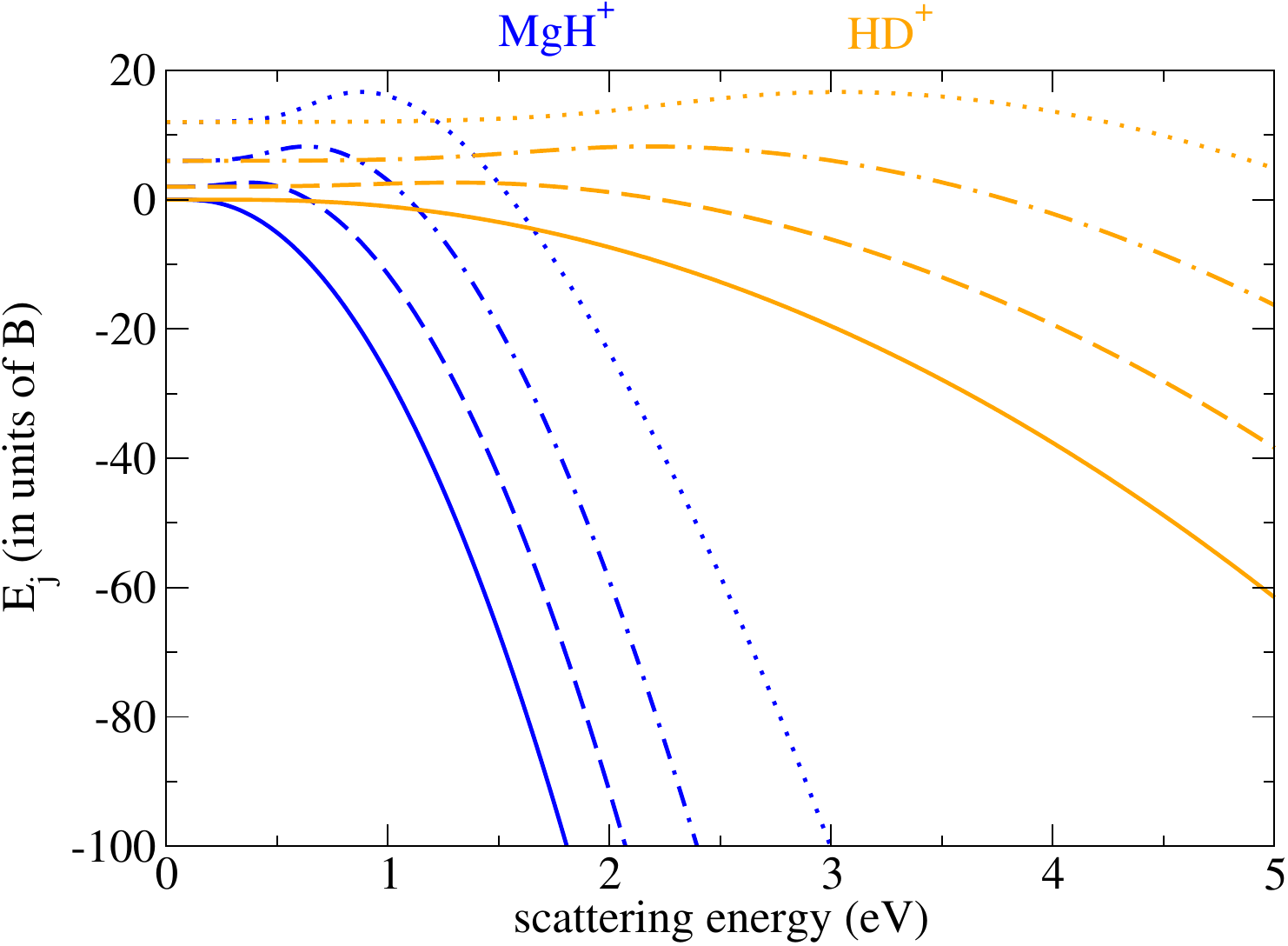}
 \caption{(Color online) Eigenvalues (in units of the rotational constant $B$)
 of the Hamiltonian, Eq.~\eqref{eq:ham1} as a function of the scattering energy for zero impact parameter. Solid, dashed, dashed dotted and dotted lines represent the states $j=0$, $j=1$ , $j=2$ , $j=3$ respectively. The field strength is calculated at $t = \frac{\tau}{2\sqrt{3}}$, where the rate of change of the Hamiltonian is largest and the probability for non-adiabatic transitions is highest.}
 \label{fig:EJ_Bnorm}
\end{figure}
Guided by the effect of the scattering energy, i.e., the electric field due to the atomic coolant on the rotational eigenstates, we identify the high-$\chi_D$ (harmonic) and low-$\chi_D$ (2-level) limits, see Fig.~\ref{fig:EJ_Bnorm}. Here MgH$^+$ represents the former limit, since its rotational eigenvalues change readily with the scattering energy, whereas HD$^+$ represents the latter limit, with its rotational eigenenergies only slightly affected by the electric field. 
\subsubsection{Low-field limit}
We begin by considering the low-coupling limit, where we use the 2-level approximation, and let $\Delta E$, the eigenstates and the matrix element $\Braket{\psi_{\iota'}(t)|\partial_t\Op{H}|\psi_0(t)}$ be constants. We obtain
\begin{equation}\label{eq:clow2}
 c(\chi_D, \kappa) = i\frac{\pi}{2\sqrt{3}}\frac{\chi_D\kappa}{\sqrt{1 + \frac{\chi_D^2}{3}}}
 \exp{\left(-\kappa\sqrt{1 + \frac{\chi_D^2}{3}}\right)}.
\end{equation}
Arriving at this expression we have made the approximations $\gamma(t) = 0$ and we have evaluated the eigen-energies at maximum field strength. 

\change{Recalling that the maximum field strength scales as $1/b^2$, the low-field expression, Eq.~\eqref{eq:clow2},} is expected to be relevant for all species and scattering energies for sufficiently large values of $b$. \change{Relating $\chi_D$ to $\varepsilon_0$ by use of Eq.~\eqref{eq:chi}, the excitation probability, i.e., the absolute square of Eq.~\eqref{eq:clow2}, to leading order is proportional to $b^{-4}$. Moreover, we}
can use our model to predict at which value of the impact parameter, $b$, maximum excitation occurs. 
The result is
\begin{equation}\label{eq:bstar}
 b \approx \sqrt{\left(\sqrt{\frac{D}{B}\sqrt{\frac{\kappa}{3}}} - \frac{\change{e^2}}{2E}\right)^2 - \left(\frac{\change{e^2}}{2E}\right)^2},
\end{equation}
with the details presented in Appendix~\ref{sec:Alow}.

It is of relevance to estimate when the adiabatic picture is applicable. Typically that is when the rotational time is short compared to the duration of the field, i.e. when $\kappa > 1$ (or even $ \gg 1$). This estimate can only be reliable when the internal rotational structure is not significantly altered by the field so as to leave $T_{rot} \propto B^{-1}$. This is the case for HD$^+$, as can be seen from Fig.~\ref{fig:EJ_Bnorm}. For HD$^+$ we therefore expect the adiabatic picture to be relevant only for $\kappa > 1$. It is seen from Fig.~\ref{fig:kappaStar} in the Appendix, that $\kappa = 1$ at around $E \approx 1.5$ eV for HD$^+$ and decreasing with energy. We therefore estimate that the adiabatic picture is only relevant for scattering energies below $1.5$ eV. In the figure we see that the adiabatic picture should lose its relevance at even smaller scattering energies for MgH$^+$, in contrast to what we see from numerical calculations. However, unlike HD$^+$ the internal rotational states are significantly affected by the external field due to the coolant, and the simple estimate based on $\kappa > 1$ for adiabaticity cannot be convincingly applied. 

\begin{figure}[tbp]
 \centering
 \includegraphics[scale=0.35]{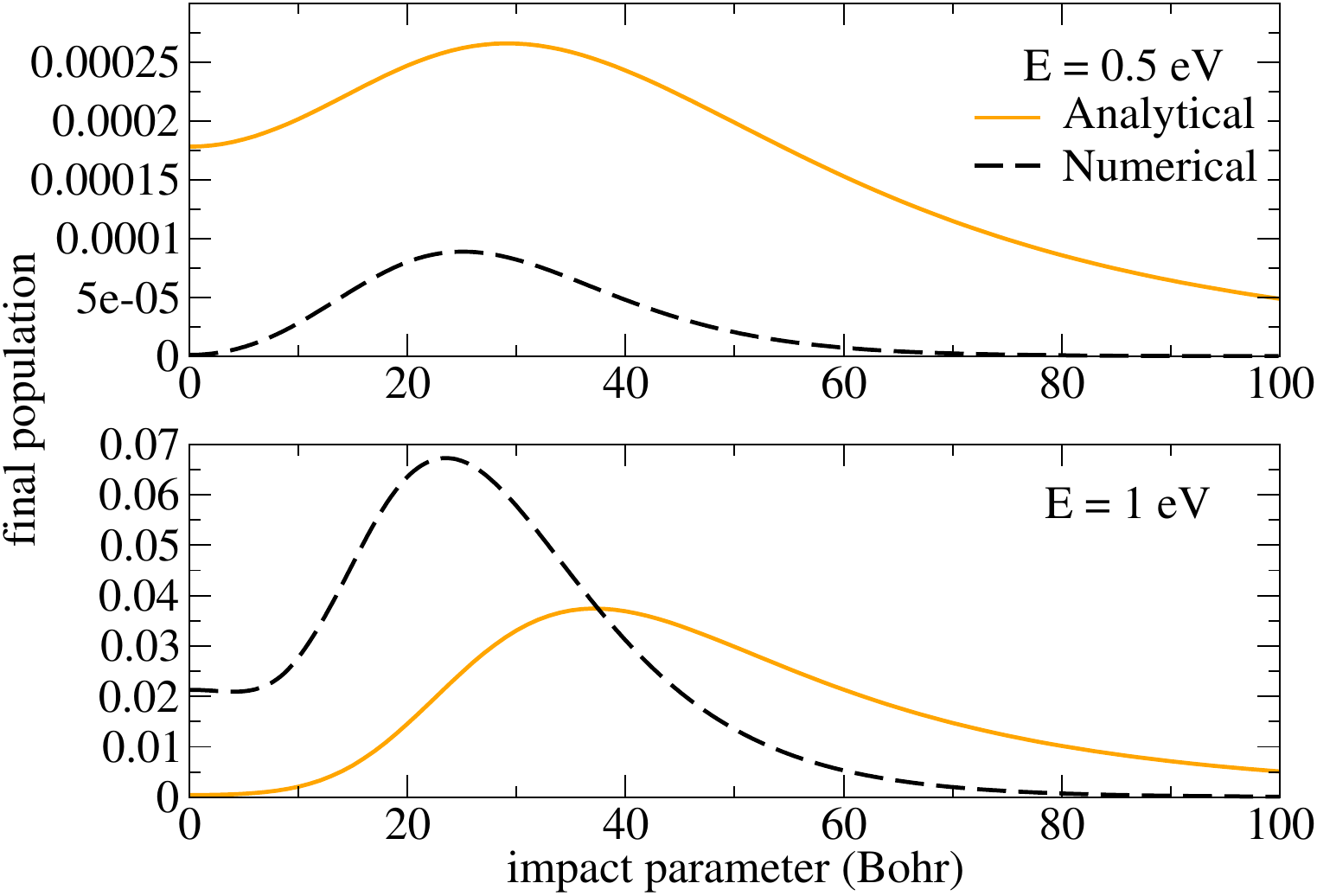}
 \caption{Population excitation of HD$^+$ as a function of impact parameter for two scattering energies, where $\kappa > 1$. The analytical results refer to the absolute square of Eq.~\eqref{eq:clow2}, and numerical to full numerical integration of Eq.~\eqref{eq:ham1}.}
 \label{fig:excHD}
\end{figure}

We now compare the population excitation obtained from the absolute square of Eq.~\eqref{eq:clow2}, i.e. the analytical formula, to the population obtained from numerical integration with the Hamiltonian Eq.~\eqref{eq:ham1}. The results are presented in Fig.~\ref{fig:excHD} for HD$^+$. We see that the analytical formula comes near at reproducing the numerical results, but the accuracy is not particularly good. In particular, for low scattering energies, upper panel, the analytical results are overestimating the numerical results. At this low scattering energy we expect both the adiabatic picture and two-level approximation to be valid, so the disagreement needs a more detailed explanation. We notice that we have completely ignored the phase $\gamma(t)$ in our expression, Eq.~\eqref{eq:clow2}, and as a result we expect our expression to overestimate the population excitation due to less cancellations due to neglect of oscillations. Furthermore, we have simply made the replacement $2B \to 2B\sqrt{1 + \frac{\chi_D^2}{3}}$ for the energy difference when evaluating the excitation. In a real scattering event the energy difference changes continually between these two extreme values. It is not straightforward to estimate whether the effect of neglecting this gradual change leads to an over- or underestimation of the resulting excitation due to the nonlinear dependence on $\chi_D$ in Eq.~\eqref{eq:clow2}. These two effects are relevant to low values of $b$, where the admixture of the excited state is more pronounced. 
For large values of $b$, the overestimation is likely due to our approximation in treating $\kappa$ as independent of $b$ (and evaluated at $b=0$). This was motivated by the observed gradual dependence on $b$. In fact, $\kappa$ is an increasing function of $b$ and due to the exponential $e^{-a\kappa}$, $a$ a constant, dependence on $\kappa$ we conclude that our model will overestimate the excitation at large $b$. As the energy increases the agreement starts to become less convincingly. Since at $1$ eV we estimate $\kappa \approx 2$ it is not likely that the disagreement is due to loss of adiabaticity, but more likely a consequence of the lack of accuracy of the 2-level model, see Fig.~\ref{fig:DEJ}.

Guided by these results we suggest a strategy on how to estimate the population excitation in the low field limit:
\begin{enumerate}
 \item Diagonalize the Hamiltonian, Eq.~\eqref{eq:ham1}, over the relevant scattering energies to obtain the rotational eigenvalues as a function of scattering energies. If the eigenenergies do not alter significantly over the relevant interval, we are in the low-field limit. (Alternatively one can say that if $\chi_D$ is not much larger than one, then we are in this limit).  
 \item If we are in the low field limit, we must also verify that the adiabatic picture is relevant. To this end we use Eq.~\eqref{eq:chistar} to estimate when $\kappa > 1$. If so, the model can give physically relevant results. 
 \item If both approximations are justified, we use the absolute square of Eq.~\eqref{eq:clow2} to estimate the excitation.
\end{enumerate}

\subsubsection{High-field limit}
In the opposite limit, i.e., high field limit, the rotational states of the molecular ions are significantly affected during scattering, we cannot find an analytical estimate for the population dynamics valid for the entire scattering event. We notice that this limit also corresponds to the harmonic limit. In this limit we find that Eq.~\eqref{eq:aTheta0}, the cosine transition moment, goes into $\frac{\theta_0^2}{2} = \sqrt{\frac{1}{2\chi_D}}$, which explains why scattering with a high $\chi_D$-value tend to lead to larger maximal orientation, $\Braket{\cos^2{\theta}}$, as seen in Fig.~\ref{fig:polarmatrix}. In this limit we can also calculate the expansion coefficient
\begin{widetext}
\begin{equation}\label{eq:cintH}
 c_1^1(t) = -\frac{1}{2}\int_{t_0}^t\frac{t}{t^{\prime 2} + \left(\frac{\tau}{2}\right)^2}\exp{\left(i3\omega_H\frac{\tau}{2}
 \ln{\left(\frac{t^\prime + \sqrt{t^{\prime 2} + \left(\frac{\tau}{2}\right)^2}}{t_0 + \sqrt{t_0^2 + \left(\frac{\tau}{2}\right)^2}}\right)}\right)}\,d t^\prime,
\end{equation}
\end{widetext}
see Section~\ref{sec:Aharm}. Here $t_0$ is an arbitrary starting time, and $\omega_H = \sqrt{2D\varepsilon_0B}$ and we remind that $\tau = 1.86\sqrt{\frac{\mu}{E^3}}$. Therefore, the phase factor in Eq.~\eqref{eq:cintH} is $\propto \sqrt{DB\mu}$ at a given scattering energy. A larger phase factor, $\omega_H\tau$, leads to faster oscillations, and thereby more cancellation of population transfer. So, Eq.~\eqref{eq:cintH} is in qualitative agreement with the plots in Fig.~\ref{fig:poplandscape}. The faster oscillations in the right panel are also consistent with the larger phase factor associated with larger values of $D$. For head-on collisions, $\omega_H\tau \propto \frac{1}{\sqrt{E}}$, and therefore we expect to find more excitations for higher energies, which is consistent with the larger oscillations in $\Braket{\cos^2{\theta}}$, corresponding to more excitations, as seen in Fig.~\ref{fig:polarmatrix}.

Although Eq.~\eqref{eq:cintH} is in qualitative agreement with the numerical results as just discussed, it dose not converge with respect to $t_0$ and we can therefore not obtain quantitative results from it. A way to understand the lack of convergence is to realize that, in a scattering event, we go from the low-field limit over moderate field strengths before reaching the high field limit. The failure to converge points to important effects occurring before reaching this limit. In addition, for large impact parameters the scattering never reaches the high-field limit, and we cannot hope to use the results obtained at this limit.
What we can do based on the high-field limit is to predict an ordering of the excitation in terms of $DB\mu$, giving less excitation for high values of the product, consistent with the values in Table~\ref{table:molpar} and the excitations in Fig.~\ref{fig:poplandscape}.
 
\begin{figure}[tbp]
 \includegraphics[scale=0.3]{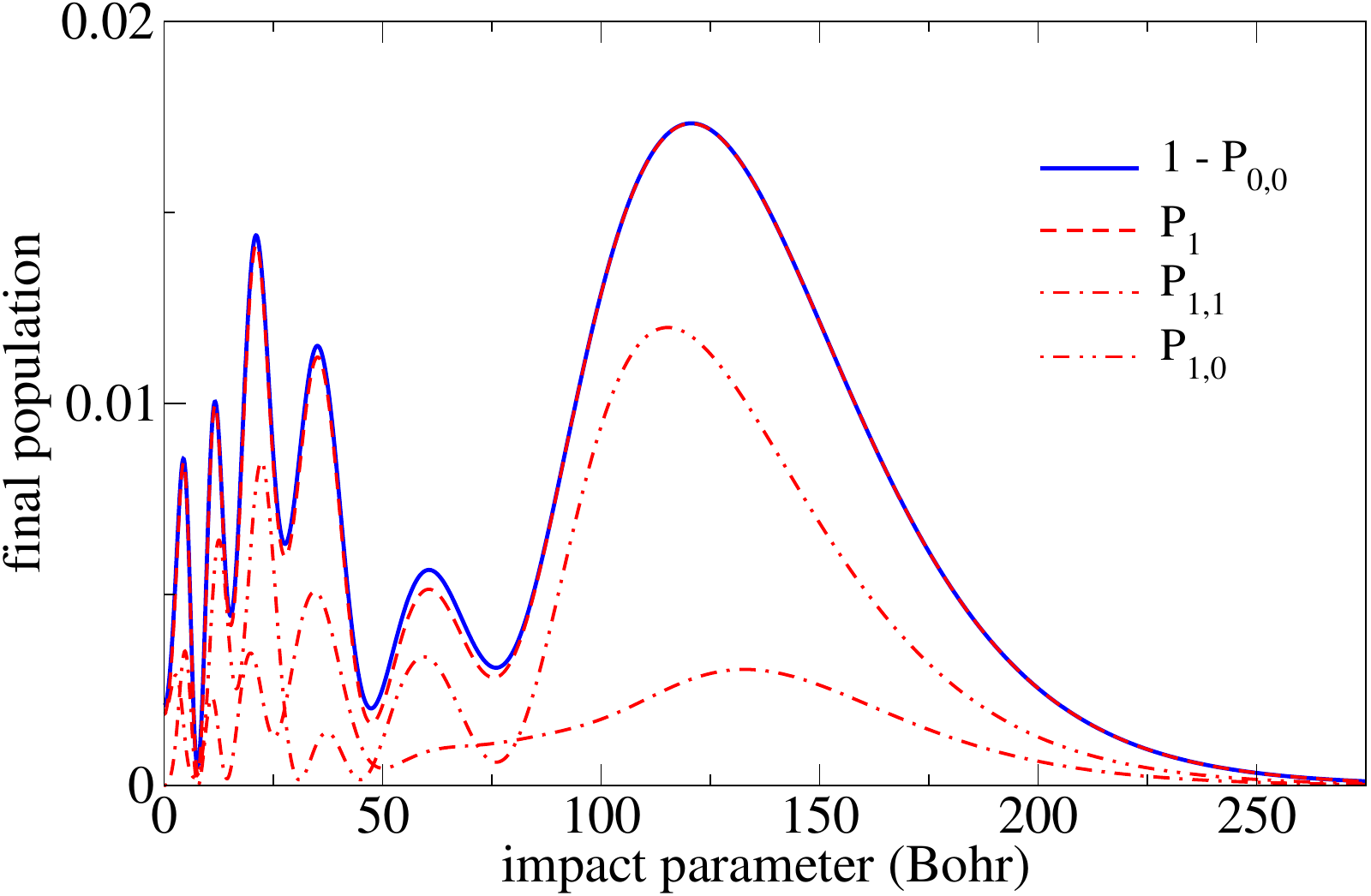}
 \caption{(Color online) Population transfer from numerical integration of the Schr\"odinger equation under the Hamiltonian Eq.~\eqref{eq:ham1} as function of impact parameter, $b$ for MgH$^+$ / Mg$^+$ at scattering energy $E=2$~eV. Total excitation (solid lines) as well as excitation to the $j=1$ sub-levels (red).}
 \label{fig:popbPolar}
\end{figure}

Finally, we present numerical results for the final population for MgH$^+$, in the high field limit in Fig.~\ref{fig:popbPolar}. We see that the final population is qualitatively the same as for the low field scattering for the low-field limit for high impact parameters. Indeed, at $b \approx 120$, where maximal excitation occurs, $\chi_D \approx 3.6$, i.e., we are closer to the low-field limit in this case. At lower impact parameters, we observe more rapid oscillations that can be understood from different values of $\omega_H$ for different impact parameters.
\begin{table}
 \begin{center}
  \begin{tabular} {c c c c c c}
  Molecule & $B$ ($10^{-5}$ H) & $D$ (at. u.) & $\mu$ ($m_e$)& $\frac{D}{B}$ ($10^{5}$) & $\sqrt{BD\mu}$\\
  \hline
  MgH$^+$ & $2.28$ & $1.18$ & $\approx 22500$ & $\approx 0.5$ & $\approx 0.78$\\ 
  MgD$^+$ & $1.5$ & $1.18$ & $\approx 22500$ & $\approx 0.8$ & $\approx 0.63$\\
  MgT$^+$ & $1.04$ & $1.18$ & $\approx 22500$ & $\approx 1.1$ & $\approx 0.53$\\
  CaH$^+$ & $2.15$ & $2.35$ & $\approx 30000$ & $\approx 1.1$ & $\approx 1.07$
  \end{tabular}
 \end{center}
 \caption{Molecular parameters of molecular ions whose rotational states are significantly affected by the electric field. Note that $\mu$ depends on the mass of the coolant as well as the mass of the molecular ion.}
 \label{table:molpar}
\end{table}

\section{Conclusions and outlook}\label{sec:Con}
In conclusion, we have studied the rotational population excitation of molecular ions in Coulomb-scattering processes. For polar molecular ions, we find that we can analyze the population excitation in the adiabatic picture. 
When the maximum interaction strength is small compared to the rotational kinetic energy, we obtain an analytical estimate of the population excitation, where the adiabatic states are obtained from the two lowest field-free eigenstates. In the opposite limit, when the (maximum) interaction energy is large compared to the rotational kinetic energy, the Hamiltonian approaches a harmonic oscillator that generate small librations around the field line resulting in a large suppression of the population excitation. In this limit we obtain an ordering of the population excitation in terms of the molecular and scattering parameters $DB\mu$ and the scattering energy. In order to achieve any detailed result in the high $\chi_D$ limit a full numerical quantum mechanical calculation is required. 

For apolar molecular ions on the other hand validity of PT has allowed us to derive an accurate closed-form estimate of the population excitation which solely depends on the molecular parameters and initial scattering energy. For a wide range of apolar molecular ions, we find the internal state to be preserved for scattering energies of 1$\,$eV and above, eventually limited by close-encounter interactions disregarded in the present treatment. Due to the symmetry associated with rotations of identical nuclei the rotational spectrum contains either only even of only odd rotational states. Since the energy difference between the 'ground' state $j=1$ and the first excited state $j=3$ is larger for the odd case we expect less energy transfer at any given energy for the odd rotational states.  

\change{Fixing the scattering energy allows for obtaining a direct relation between population excitation and molecular parameters
(rotational constants and coupling parameters, i.e., the molecular polarizability, quadrupole moment or permanent dipole moment). Our study thus suggests that collisions can be utilized as a spectroscopic tool to measure these molecular parameters.
Furthermore, }
rotational excitation of the molecular ion as a consequence of collisions with atomic ions has implications for sympathetic cooling. Since sympathetic cooling consists of a series of such collisions, it is prone to accumulated rotational population redistribution which would be detrimental to most envisioned applications of the molecular ions. Our results here form the basis for investigating how rotational excitations accumulate over a full cooling cycle, as reported in the companion paper~\cite{BerglundPRA2}. 

\begin{acknowledgments}
We would like to thank Stefan Willitsch for fruitful discussions on the quadrupole aspect of the work. 
Financial support from the State Hessen Initiative for the Development of Scientific and Economic Excellence (LOEWE), the European Commission’s FET Open TEQ, the Villum Foundation, and the Independent
Research Fund Denmark is gratefully acknowledged.
This research was supported in part by the National Science Foundation under Grant No. NSF PHY-1748958.
\change{JMB wishes to thank SECIHTI for providing a postdoc scholarship.}
\end{acknowledgments}

\appendix
\section{Adiabatic picture}\label{sec:AAdia}
\subsection{Neglect of vibrational excitations}\label{sec:Avib}
Vibrational excitations may occur when the dipole moment changes with bond length. Upon Taylor expanding the dipole moment around the equilibrium distance the first term is $\left.\frac{dD}{dx}\right|_0x$, with $x$ the deviation from equilibrium. In the harmonic approximation the transition moment of $x$ between the vibrational states $\nu = 0 \to \nu = 1$ is $\sqrt{\frac{1}{2\mu_{vib}\omega_0}}$~\cite{atkins2011molecular}. The difference in energy between the two states is given by $\Delta E = \omega_0$, where $\omega_0$ is the vibrational frequency. Therefore $\chi_{vib} = \left.\frac{dD}{dx}\right|_0\varepsilon_0\sqrt{\frac{1}{2\mu_{vib}\omega_0^3}}
= \left.\frac{dD}{dx}\right|_0\sqrt{\frac{1}{2\mu_{vib}\omega_0^3}}E^2$. We estimate that for MgH$^+$ at $E=2$ eV $\chi_{\nu} \approx 8 \cdot 10^{-2}$. 
We will therefore neglect vibrational excitations and only consider the rotational degree of freedom in our model.
\subsection{Obtaining the expansion coefficients}\label{sec:Aexp}
Consider the instantaneous eigenstates $\Ket{\psi_n(t)}$
of a time-dependent Hamiltonian $\Op{H}(t)$ of the form \eqref{eq:ham1}, 
\begin{equation}\label{eq:AAd1}
 \Op{H}(t)\Ket{\psi_{\iota}(t)} = E_{\iota}(t)\Ket{\psi_{\iota}(t)},
\end{equation}
with instantaneous eigenenergies $E_{\iota}(t)$.
Any state $\Ket{\Psi(t)}$ can be expanded into the time-dependent eigenstates, $\Ket{\Psi(t)} = \sum_{\iota} c_{\iota}e^{i\Theta_{\iota}(t)}\Ket{\psi_{\iota}(t)}$, where $\Theta_{\iota} = -\int^tE_{\iota}(t')dt'$.
Inserting the expansion of $\Ket{\Psi(t)}$ into the time-dependent Schr\"odinger equation,
\begin{equation}
 i\partial_t\Ket{\psi(t)} = \Op{H}\Ket{\psi(t)}\,,
\end{equation}
and multiplying both sides by $\Bra{\psi_{\iota'}(t)}$, we obtain 
\begin{widetext}
 \begin{equation}\label{eq:Aeom_ci}
  \dot{c}_{\iota'}(t) = -c_{\iota'}(t)\Braket{\psi_{\iota'}(t)|\dot{\psi}_{\iota'}(t)} -\sum_{\iota \neq  \iota'} c_{\iota}(t)e^{i\Delta\Theta_{\iota\iota'}(t)}\Braket{\psi_{\iota'}(t)|\dot{\psi}_{\iota}(t)}.
 \end{equation}
\end{widetext}
Differentiating Eq.~\eqref{eq:AAd1} w.r.t. time,  we find after multiplying with $\Bra{\psi_{\iota'}(t)}$ from the left
\begin{equation}
 \Braket{\psi_{\iota'}(t)|\dot{\psi}_{\iota}(t)} = \frac{\Braket{\psi_{\iota'}(t)|\partial_t\Op{H}(t)|\psi_{\iota}(t)}}{E_{\iota}(t)-E_{\iota'}(t)}, \qquad \iota \neq \iota',
\end{equation}
where we have used $\Braket{\psi_{\iota'}(t)|\psi_{\iota}(t)} = \delta_{\iota'\iota}$.
Assuming a nearly adiabatic dynamics,    
$|c_{\iota = 0}(t)| \sim 1$ and $|c_{\iota}(t)| \ll 1, \iota \neq 0$, we can then write $c_{\iota = 0}(t) \sim e^i\gamma_0(t)$, where $\gamma_0(t)$ is the Berry phase~\cite{sakurai1995modern}.

we can integrate Eq.~\eqref{eq:Aeom_ci},
\begin{equation}\label{eq:Acadia}
  c_{\iota'}(t) \approx \int_{-\infty}^t e^{i\gamma_0(t^\prime)} e^{i\Delta\Theta_{0\iota'}(t^\prime)}\frac{\Braket{\psi_{\iota'}(t^\prime)|\partial_t\Op{H}(t^\prime)|\psi_0(t^\prime)}}{E_{\iota'}(t^\prime)-E_0(t^\prime)}dt^\prime.
\end{equation}
\subsubsection{Evaluation of the expansion coefficient in the low-field limit}\label{sec:Alow}
In order to evaluate the coefficient, Eq.~\eqref{eq:Acadia} we make the approximation that $\Delta\Theta_{0\iota'}(t) \approx \Delta E_{\iota'0}t$, where $\Delta E_{\iota'0}$ is constant. In this approximation the state $\Ket{\psi_0(t)}$ is going to be approximately the field-free ground state, and consequently the Berry phase is approximately zero~\cite{sakurai1995modern}. This approximation is expected to work better for low interaction scattering, i.e. low $\chi_D$, since the effect on the eigenenergies increases with $\chi_D$. Within this approximation we may then write
\begin{widetext}
\begin{equation}
 c_{\iota'}(t) \approx \frac{D\varepsilon_0}{\Delta E_{\iota'0}}\left(\frac{\tau}{2}\right)^2\int_{-\infty}^{\infty} e^{i\Delta E_{\iota'0}t}\frac{2tdt}{\left(t^2+\left(\frac{\tau}{2}\right)^2\right)^2}\Braket{\psi_{\iota'}|\cos{\theta_a}|\psi_0}.
\end{equation}
\end{widetext}
Making use of the Fourier relation between the transform of a function and its derivative
\begin{widetext}
\begin{equation}
 c_{\iota'}(t) \approx iD\varepsilon_0\left(\frac{\tau}{2}\right)^2\int_{-\infty}^{\infty} \frac{e^{i\Delta E_{\iota'0}t}dt}{\left(t+i\frac{\tau}{2}\right)\left(t-i\frac{\tau}{2}\right)}\Braket{\psi_{\iota'}|\cos{\theta_a}|\psi_0},
\end{equation}
\end{widetext}
we obtain an expression that can be solved by Cauchy's integral formula
\begin{equation}
 c_{\iota'}(t) \approx i\frac{\pi}{2}D\varepsilon_0\tau e^{-\frac{\Delta E \tau}{2}}\Braket{\psi_{\iota'}|\cos{\theta_a}|\psi_0}.
\end{equation}
In the low field limit we can use the field free states, i.e. $\iota = 0 \leftrightarrow j=0$. In this limit we have $\Delta E_{\iota 0} = 2B$ and $\Braket{\cos{\theta_a}} = \frac{1}{\sqrt{3}}$. The expansion coefficient is then
\begin{equation}\label{eq:AcPert1}
 c_{\iota'}(t) \approx i\frac{\pi}{2\sqrt{3}}\chi_D\kappa e^{-\kappa}.
\end{equation}

As an improvement to the field-free limit we consider the perturbative limit, in which we let the field-dressed states be described by a superposition of a minimal free-field basis. In this limit the lowest eigenvalues are $E_{\pm} = B\left(1 \pm \sqrt{1 + \frac{\chi_D^2}{3}}\right)$, and consequently $\Delta E = 2B\sqrt{1 + \frac{\chi_D^2}{3}}$. Therefore, letting $2B \to 2B\sqrt{1 + \frac{\chi_D^2}{3}}$ in Eq.~\eqref{eq:AcPert1} we have
\begin{equation}\label{eq:cPert2}
\begin{aligned}
 c_{\iota = 2}(t) &\approx i \frac{\pi}{2}\chi_D\kappa e^{-\kappa\sqrt{1+\frac{1}{3}\chi_D^2}}
  \Braket{\psi_{\iota = 2}|\cos{\theta_a}|\psi_{\iota=0}} \\
  &= i \frac{\pi}{2\sqrt{3}}\frac{\chi_D\kappa}{\sqrt{1 + \frac{\chi_D^2}{3}}} e^{-\kappa\sqrt{1+\frac{1}{3}\chi_D^2}}.
  \end{aligned}
\end{equation}
We will now proceed to calculate the transition moment.

Consider the Hamiltonian Eq.~\eqref{eq:ham1}. A minimal basis representation is given by the states $\Ket{0,0}$ and $\Ket{1,0}$, in which we can represent the Hamiltonian as
\begin{equation}\label{eq:HamMin}
 \Op{H} \to 
 \begin{pmatrix}
  0 & V  \\
  V & 2B
 \end{pmatrix}.
\end{equation}
Here $V = -\frac{D\varepsilon}{\sqrt{3}}$. The eigenvectors of the Hamiltonian Eq.~\eqref{eq:HamMin} are
\begin{subequations}
 \begin{equation}
  \psi_0 = \frac{1}{\sqrt{E_1^2 + V^2}}
  \begin{pmatrix}
  E_2 \\
  -V
  \end{pmatrix}
 \end{equation}
 and
 \begin{equation}
  \psi_1 = \frac{1}{\sqrt{E_0^2 + V^2}}
  \begin{pmatrix}
  E_1 \\
  -V
  \end{pmatrix}
 \end{equation}
\end{subequations}
with the corresponding eigenvalues
\begin{subequations}\label{eq:eigenenergies}
 \begin{equation}
  E_0 = B - \sqrt{B^2 + V^2}
 \end{equation}
 \begin{equation}
  E_1 = B + \sqrt{B^2 + V^2},
 \end{equation}
\end{subequations}
and consequently $\Delta E = 2\sqrt{B^2 + V^2} = 2B\sqrt{1 + \frac{\chi_D^2}{3}}$.
Let's now evaluate some combinations of $E_0$ and $E_1$ that follow from Eq.~\eqref{eq:eigenenergies}. The sum of the eigenenergies is easily evaluated to be
$E_0 + E_1 = 2B$, $E_0E_1 = -V^2$ and $E_0^2 + E_1^2 = 2\left(2B^2 + V^2\right)$.
The representation of the transition operator in the minimal field free basis is
 \begin{equation}
  \cos{\theta_a} = \frac{1}{\sqrt{3}}
  \begin{pmatrix}
   0 & 1 \\
   1 & 0
  \end{pmatrix}.
 \end{equation}
Now let's consider the transition moment, 
\begin{widetext}
 \begin{equation}
 \begin{aligned}
   \Braket{\psi_1|\cos{\theta_a}|\psi_0} &= \frac{1}{\sqrt{\left(E_0^2 + V^2\right)\left(E_1^2+V^2\right)}}
   \begin{pmatrix}
    E_0, & V
   \end{pmatrix}
   \frac{1}{\sqrt{3}}
   \begin{pmatrix}
    0 & 1 \\
    1 & 0
   \end{pmatrix}
   \begin{pmatrix}
    E_1 \\
    V
   \end{pmatrix}
   = \frac{V\left(E_0 + E_1\right)}{\sqrt{3}\sqrt{\left(E_0^2 + V^2\right)\left(E_1^2+V^2\right)}} \\
   &= \frac{2BV}{\sqrt{3}\sqrt{4V^2}\sqrt{B^2 + V^2}} 
   = \frac{1}{\sqrt{3}\sqrt{1 + \frac{\chi_D^2}{3}}},
   \end{aligned}
\end{equation}
\end{widetext}
which is our final result in the low-field limit.

\subsubsection{High field limit of the polar Hamiltonian}\label{sec:Aharm}
In the extreme of high fields the molecule is forced to align with the external field. The remaining kinetic energy fights the field and the molecule describes so called 
librating motion, i.e.~restrained rotation. The trigonometric functions in the Laplacian and in the interaction term can be approximated by
$\cos{\theta} \approx 1 - \frac{\theta^2}{2}$ and $\sin{\theta} \approx \theta$. The Hamiltonian is now approximately
\begin{equation}
 \Op{H} = -B\left(\frac{1}{\theta}\frac{\partial}{\partial \theta}\left(\theta\frac{\partial}{\partial \theta}\right) + 
 \frac{1}{\theta^2}\frac{\partial^2}{\partial \phi^2}\right) + \frac{1}{2}D\varepsilon \theta^2 - D\varepsilon,
\end{equation}
i.e.~the polar angle $\theta$ takes the role of the radial coordinate in 2D-polar coordinates and the azimutal angle $\phi$ that 
of the standard polar angle.
Assuming a separation of variables $\psi(\theta,\phi) = g(\theta)e^{im\phi}$ the function $g(\theta)$ satisfies
\begin{equation}\label{eq:diff1}
 g''(\theta) + \frac{1}{\theta} g'(\theta) + \left(\tilde{E} - \frac{1}{\theta^2}m^2 - \frac{1}{2}\chi\theta^2\right)g = 0,
\end{equation}
where $\tilde{E} = \frac{E}{B} + \chi$. Considering Eq.~\eqref{eq:diff1} for large $\theta$ where terms 2 and 4 are negligible we obtain 
$g(\theta) \sim \exp{\left(-\frac{\theta^2}{2\theta_0^2}\right)}$, where
\begin{equation}\label{eq:aTheta0}
 \frac{1}{\theta_0^2} = 2\chi_D.
\end{equation}
In the opposite limit we can neglect 
terms 1 and 5 and we get $g(\theta) \sim \theta^{|m|}$. According to Sturm-Liouville theory the equation~\eqref{eq:diff1} is not in self-adjoint form, but the
integrating factor is simply $\theta$. The standard integration interval wold  be $[0,\infty]$, but we have $\theta \in [0,\pi]$. If, 
however $\theta_0 \ll \pi$ we may approximately let $\pi \to \infty$. The eigenvalues and eigenfunctions are given by
\begin{subequations}
\begin{equation}\label{eq:aWfG}
 \psi_0^0(\theta,\phi) = \frac{1}{\sqrt{\pi\theta_0^2}} \exp{\left(-\frac{\theta^2}{2\theta_0^2}\right)},
\end{equation}
\begin{equation}
 \psi_0^2(\theta,\phi) = \left(\frac{\theta^2}{\theta_0^2} - 1\right)
 \frac{\exp{\left(-\frac{\theta^2}{2\theta_0^2}\right)}}{\sqrt{\pi\theta_0^2}},
\end{equation}
\end{subequations}
with the corresponding eigenvalues
\begin{subequations}
\begin{equation}\label{eq:aE0}
 E_{0,0} + D\varepsilon  = \sqrt{2BD\varepsilon}.
\end{equation}
\begin{equation}
 E_{0,2} + D\varepsilon  = 3\sqrt{2BD\varepsilon},
\end{equation}
\end{subequations}
thereby defining the harmonic energy scale
\begin{equation}\label{eq:aOmegaH}
 \omega_H = \sqrt{2BD\varepsilon}.
\end{equation}
In order to evaluate the matrix elements we need to evaluate expressions such as
$\cos{\theta_a} \approx \left(1-\frac{\theta_a^2}{2}\right)$. Notice that
\begin{equation}
 \frac{2}{\theta_0^2}\int_0^\infty\theta^n\theta\exp{\left(-\frac{\theta_a^2}{\theta_0^2}\right)} d\theta = n!!\left(\frac{\theta_0^2}{2}\right)^{\frac{n}{2}},
\end{equation}
where $n!! = 2\cdot 4 \cdot 6 \cdot \dots \cdot n$. Then
\begin{widetext}
\begin{equation}\label{eq:cosHarm}
\begin{aligned}
  \Braket{2,0|-\frac{\theta^2}{2}|0,0} &\propto -\frac{2}{2\theta_0^2}\int_0^{\infty}\left(\frac{\theta_a^2}{\theta_0^2}-1\right)\theta_a^3\exp{\left(-\frac{\theta_a^2}{\theta_0^2}\right)} d\theta_a                                     = -\frac{1}{2}\left( \frac{4!!}{\theta_0^2}\left(\frac{\theta_0^2}{2}\right)^2 - 2!!\left(\frac{\theta_0^2}{2}\right)\right) 
 = -\frac{1}{2}\theta_0^2
  \end{aligned}
 \end{equation}
\end{widetext}

In the adiabatic limit the second term of Eq.~\eqref{eq:Aeom_ci} tends to zero and we are left with a separable differential equation for $c_{\iota=0}(t)$, which has the solution ($c_{\iota=0}(t=0) = 1$)
\begin{equation}
 c_{\iota=0}(t) = e^{i\int_{-t_0}^{t_0}\Braket{\psi_{\iota=0}(t)|\dot{\psi}_{\iota=0}(t)}\,d t} \equiv e^{i\gamma_0}.
\end{equation}
From Eqs.~\eqref{eq:aWfG} and \eqref{eq:aE0}, $\psi_0^0(\theta,\phi) = \frac{1}{\sqrt{\pi\theta_0^2}} \exp{\left(-\frac{\theta^2}{2\theta_0^2}\right)}$, with eigenvalue $E_0 = \sqrt{2BD\varepsilon}$, and therefore 
\begin{equation}\label{eq:AphaseGamma}
 \begin{aligned}
  \gamma_0(t) &= \sqrt{2D\varepsilon_0B}\frac{\tau}{2}\int_{t_0}^t\frac{\,d t'}{\sqrt{t'^2 + \left(\frac{\tau}{2}\right)^2}} \\
  &= \sqrt{2D\varepsilon_0B}\frac{\tau}{2}
 \ln{\left(\frac{t + \sqrt{t^2 + \left(\frac{\tau}{2}\right)^2}}{t_0 + \sqrt{t_0^2 + \left(\frac{\tau}{2}\right)^2}}\right)}.
 \end{aligned}
\end{equation}

The dynamical phase, $\Theta_{0,\iota'}$ is 
\begin{equation}\label{eq:AphaseTheta}
\begin{aligned}
 \Theta_{0,\iota'}(t) &= 2\sqrt{2D\varepsilon_0B}\frac{\tau}{2}\int_{t_0}^t\frac{\,d t'}{\sqrt{t'^2 + \left(\frac{\tau}{2}\right)^2}} \\
 &= 2\sqrt{2D\varepsilon_0B}\frac{\tau}{2}
 \ln{\left(\frac{t + \sqrt{t^2 + \left(\frac{\tau}{2}\right)^2}}{t_0 + \sqrt{t_0^2 + \left(\frac{\tau}{2}\right)^2}}\right)}
 \end{aligned}
\end{equation}
Combining Eqs.~\eqref{eq:AphaseGamma} and ~\eqref{eq:AphaseTheta}, we obtain the total phase
\begin{equation}\label{eq:AphaseTot}
 \gamma_0(t) + \Theta_{0,\iota'}(t) = 3\omega_H\frac{\tau}{2}
 \ln{\left(\frac{t + \sqrt{t^2 + \left(\frac{\tau}{2}\right)^2}}{t_0 + \sqrt{t_0^2 + \left(\frac{\tau}{2}\right)^2}}\right)},
\end{equation}
where we have used Eq.~\eqref{eq:aOmegaH} for the harmonic zero point energy.

The non-phase part of the derivative is 
\begin{widetext}
\begin{equation}\label{eq:Acdot}
\begin{aligned}
 \frac{\Braket{\iota'|\frac{\partial \Op{H}(t)}{\partial t}|0}}{\Delta E_{\iota',0}(t)} &=  -D\frac{d\varepsilon(t)}{dt}\frac{\Braket{\iota'|\cos{\theta_a}(t)|0}}{\Delta E_{\iota',0}(t)} =
 -2D\varepsilon_0\left(\frac{\tau}{2}\right)^2\frac{t}{\left(t^2 + \left(\frac{\tau}{2}\right)^2\right)^2}\frac{\theta_0^2}{2}
 \frac{\sqrt{t^2 + \left(\frac{\tau}{2}\right)^2}}{2\sqrt{2D\varepsilon_0B}\frac{\tau}{2}} \\
 &= -\sqrt{\frac{\chi_D}{2}}\frac{\tau}{2}\frac{t}{\left(t^2 + \left(\frac{\tau}{2}\right)^2\right)^{3/2}}\frac{\theta_0^2}{2} 
 = -\frac{1}{2}\frac{t}{\left(t^2 + \left(\frac{\tau}{2}\right)^2\right)} ,
 \end{aligned}
\end{equation}
\end{widetext}
where in the last line we have used $\theta_0^2 = \sqrt{\frac{2}{\chi_D}}\frac{\sqrt{t^2 + \left(\frac{\tau}{2}\right)^2}}{\tau/2}$. It is remarkable that this expression becomes independent on the dipole moment, rotational constant and field strength. Notice also, that our expression, the angle increases without bound as the field diminishes.
Combining Eqs.~\eqref{eq:cosHarm}, \eqref{eq:AphaseTot} and \eqref{eq:Acdot} we obtain for the expansion coefficient in the high field limit
\begin{widetext}
\begin{equation}
 c_2^0(t)  = -\frac{1}{2}\int_{t_0}^t\frac{t}{t^{\prime 2} + \left(\frac{\tau}{2}\right)^2}\exp{\left(i3\omega_H\frac{\tau}{2}
 \ln{\left(\frac{t^\prime + \sqrt{t^{\prime 2} + \left(\frac{\tau}{2}\right)^2}}{t_0 + \sqrt{t_0^2 + \left(\frac{\tau}{2}\right)^2}}\right)}\right)}\,d t^\prime.
\end{equation}
\end{widetext}

\subsection{Basis set comparison}\label{sec:Basis}
\begin{figure}[tbp]
 \centering
 \includegraphics[scale=0.3]{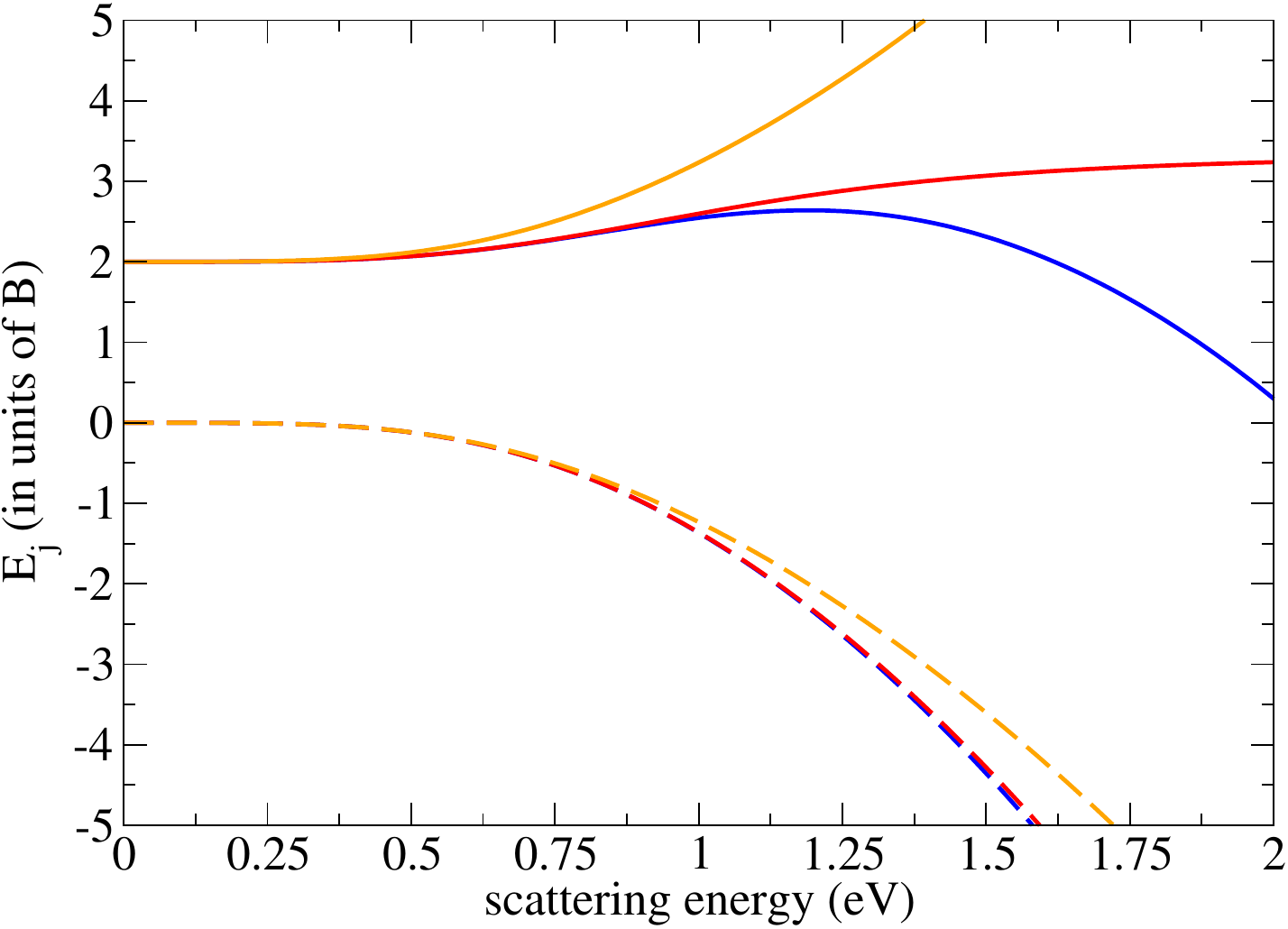}
 \hfill
  \centering
  \includegraphics[scale=0.3]{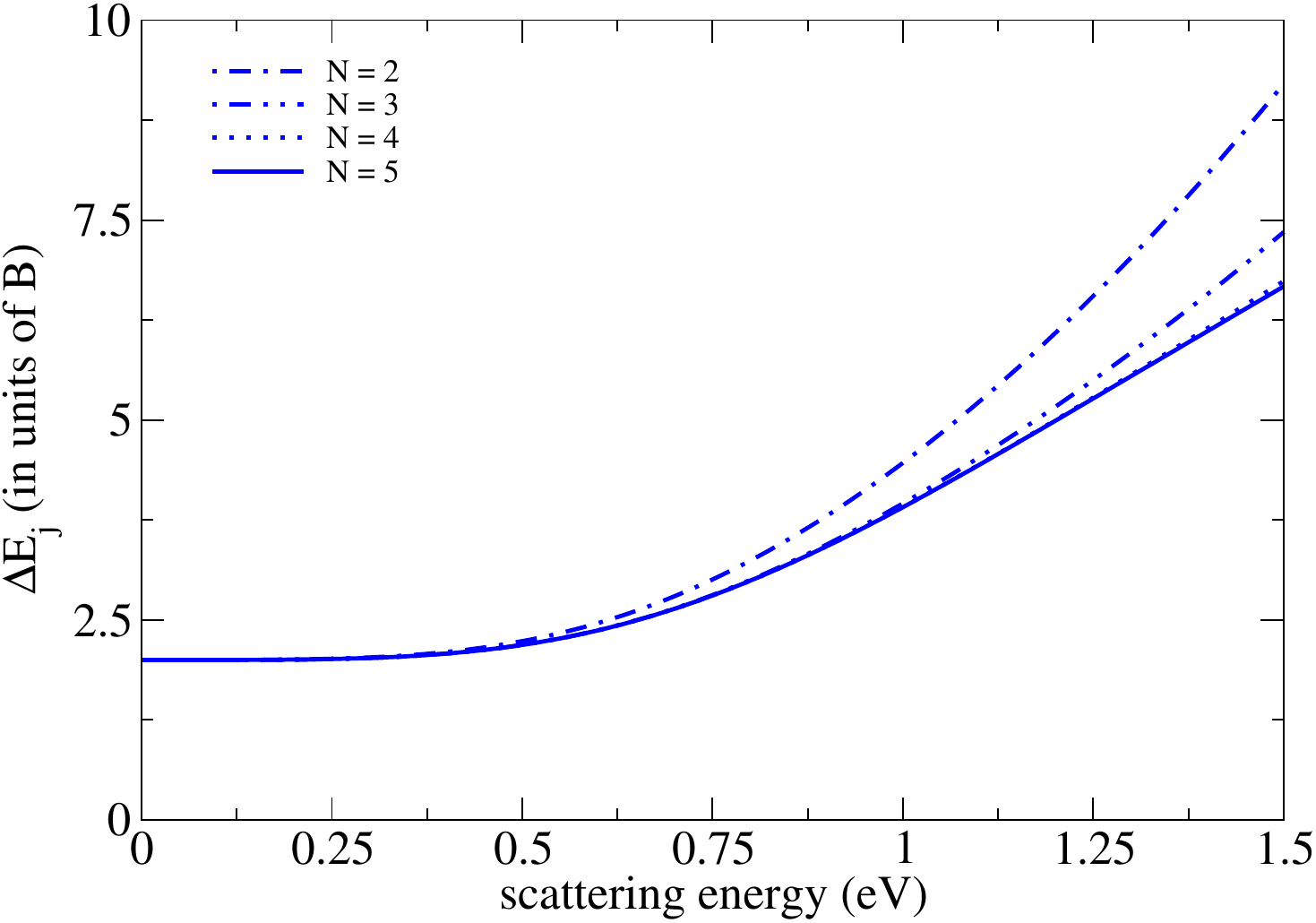}
 \caption{Comparison of the eigenenergies for different numbers of field-free states, $N$, used as basis set to represent the field-dressed states as function of scattering energy for the HD$^+$ ion. Upper panel: Rotational eigen-energies as a function of the scattering energy. Striped lines indicate the ground state $J=0$ and the solid lines indicate the state $J=1$. Blue  lines use a basis of $N=5$ free-field states, red $N=3$ and orange $N=2$. Lower panel: Energy difference between the first two field dressed states using different numbers of field-free states, ($N$), to represent them.}
 \label{fig:DEJ}
\end{figure}

We compare our analytical 2-level approximation of the expansion coefficient~\eqref{eq:clow2} (and therefore the population excitation) to numerical calculations where the eigenstates and eigenvalues are obtained from numerical diagonalization of the Hamiltonian~\eqref{eq:ham1}. Looking at the energy differences as a function of scattering energy for different sizes of the field-free states used to form the field-dressed states we see that two field-free states ($N=2$) are not sufficient to reproduce the converged energy difference, $N=5$, see Fig.~\ref{fig:DEJ}. We see that the energy difference is overestimated by $N=2$. We notice that already at $N=3$ can the energy difference be approximately achieved.  

\subsection{Evaluation of $f(\kappa)$}\label{sec:Aint}
\begin{figure}[tbp]
 \centering
 \includegraphics[scale=0.325]{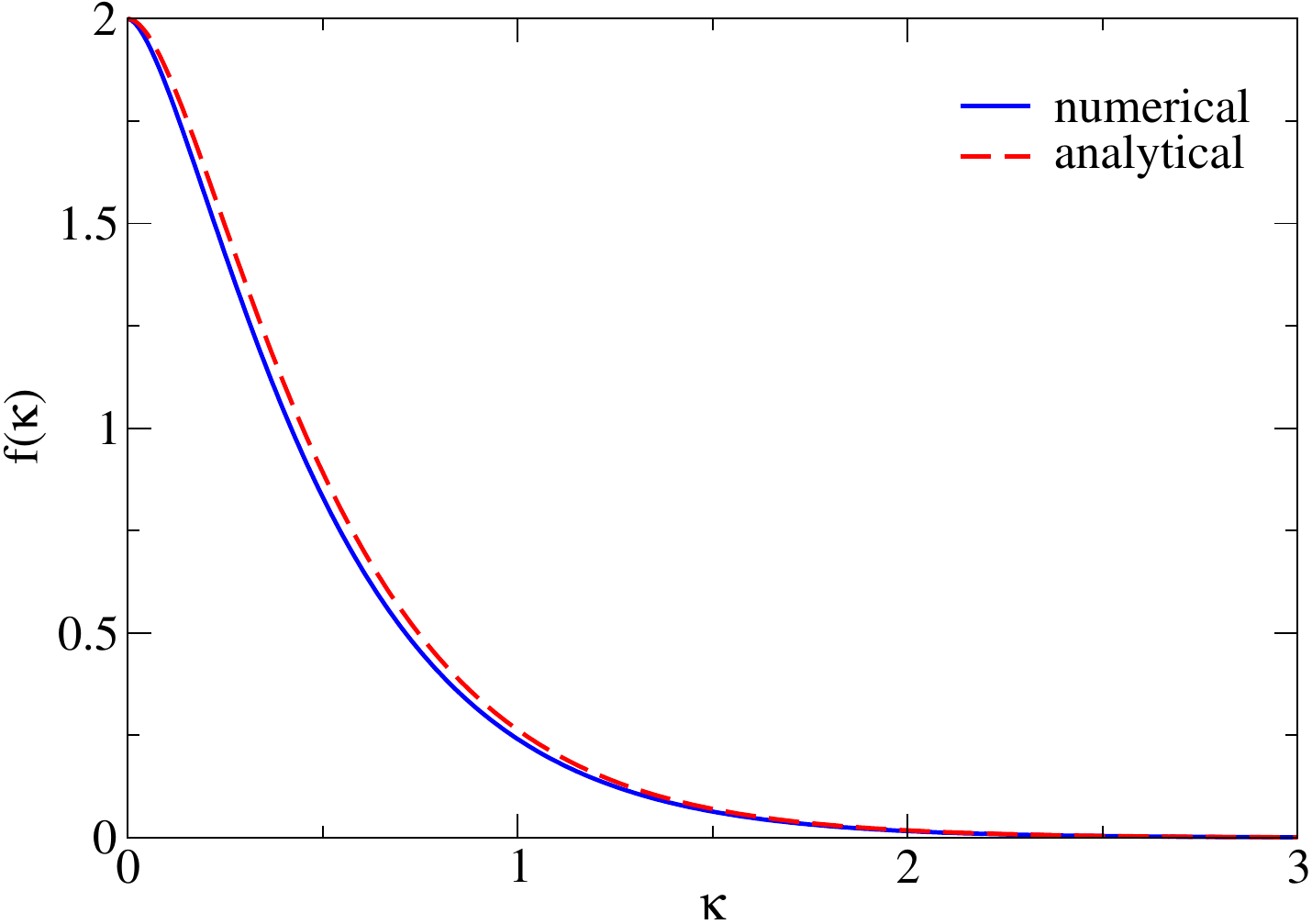}
 \caption{Comparison of the numerical integration of the function $f(\kappa) = \int_{-\pi/2}^{\pi/2} \cos{u}e^{3i\kappa\tan{u}}du$, Eq.~\eqref{eq:Afkappa} and the analytical approximation $f(\kappa) = 2\sqrt{1+6\kappa}e^{-3\kappa}$, Eq.~\eqref{eq:Afkappafin}.}
 \label{fig:AfkappaComp}
\end{figure}

We wish to evaluate the integral 
\begin{equation}\label{eq:Afkappa}
 f(\kappa) = \int_{-\pi/2}^{\pi/2}\cos{u}e^{i3\kappa\tan{u}}du.
\end{equation}
First notice that
\begin{equation}\label{eq:Af0}
 f(0) = \int_{-\pi/2}^{\pi/2}\cos{u}du = 2.
\end{equation}
We continue by considering the derivative of $f(\kappa)$ with respect to $\kappa$ and make use of Leibniz' rule for differentiating under the integral
\begin{equation}\label{eq:Adfkappadkappa}
 \begin{aligned}
  f'(\kappa) &= 3i\int_{-\pi/2}^{\pi/2}\cos{u}\tan{u}e^{i3\kappa\tan{u}}du \\
  &= 3i\int_{-\pi/2}^{\pi/2}\sin{u}e^{i3\kappa\tan{u}}du.
 \end{aligned}
\end{equation}
Next we integrate by parts and since $\left.\cos{u}e^{3i\kappa\tan{u}}\right|_{-\pi/2}^{\pi/2} = 0$ we get
\begin{equation}\label{eq:Adfkappadkappa2}
\begin{aligned}
 f'(\kappa) &= -9\kappa\int_{-\pi/2}^{\pi/2}\cos{u}(1+\tan^2{u})e^{i3\kappa\tan{u}}du \\
 &= -9\kappa f(\kappa) -9\kappa \frac{1}{(3i)^2}\frac{d^2f(\kappa)}{\,d\kappa^2} \\
 &=\kappa\frac{d^2f(\kappa)}{d\kappa^2} -9\kappa f(\kappa),
 \end{aligned}
\end{equation}
where in the second line we have twice made use of $\frac{1}{3i}\frac{d}{d\kappa}e^{i3\kappa\tan{u}} = \tan{u}e^{i3\kappa\tan{u}}$.
So, $f(\kappa)$ satisfies the differential equation
\begin{equation}\label{eq:diffkappa}
 f''(\kappa) - \frac{1}{\kappa}f'(\kappa) - 9f(\kappa) = 0.
\end{equation}
Notice that for large $\kappa$ the middle term is suppressed by the $\frac{1}{\kappa}$ dependence, and hence in this limit
\begin{equation}
 f''(\kappa) - 9f(\kappa) = 0
\end{equation}
with the solutions $f(\kappa) = Ae^{3\kappa} + Be^{-3\kappa}$. Since $\kappa \ge 0$ and we are looking for square integrable functions that are finite for $\kappa \to \infty$ we must require $A = 0$. Our ansatz is therefore to look for $f(\kappa) = Bg(\kappa)e^{-3\kappa}$, where $g(\kappa)$ is to be found. Using our ansatz in Eq.~\eqref{eq:diffkappa} we arrive at the differential equation for $g(\kappa)$
\begin{equation}
 \kappa g''(\kappa) - \left(1+6\kappa\right)g'(\kappa) + 3g(\kappa) = 0.
\end{equation}
We make the further ansatz for $g(\kappa)$
\begin{equation}
 \begin{cases}
  g(\kappa) = \sqrt{1+6\kappa} \\
  g'(\kappa) = \frac{3}{\sqrt{1+6\kappa}} \\
  g''(\kappa) = -\frac{9}{\sqrt{1+6\kappa}^3}
 \end{cases}
\end{equation}
from which
\begin{equation}
 3g(\kappa)  - \left(1+6\kappa\right)g'(\kappa) = 0
\end{equation}
follows. We have therefore approximately solved the integral as long as the second derivative term $-\frac{9\kappa}{\sqrt{1+6\kappa}^3} \approx 0$, which becomes exact when $\kappa \to 0, \infty$. With this functional form and Eq.~\eqref{eq:Af0} we obtain $B = 2$, with the final result
\begin{equation}\label{eq:Afkappafin}
 f(\kappa) \approx 2\sqrt{1+6\kappa}e^{-3\kappa}.
\end{equation}
The analytical expression, Eq.~\eqref{eq:Afkappafin} and the exact form, Eq.~\eqref{eq:Afkappa} are plotted in Fig.~\ref{fig:AfkappaComp}.

\subsection{Molecular model}
For completeness, we present the Hamiltonian for apolar molecular ions,
Eq.~\eqref{eq:ham2} in the main text, with $\Op\theta_a$ substituted by $\beta$, $\Op\theta$ and $\Op\phi$.
It reads
\begin{widetext}
\begin{eqnarray}
\nonumber
\Op{H}_{ap} &=&  B\Op{J}^2 - \frac{\varepsilon^2(t)}{4}\Big[\Delta\alpha\Big(\cos^2{\beta}\cos^2{\Op{\theta}} 
+2\cos{\beta}\sin{\beta}\cos{\Op{\theta}}\sin{\Op{\theta}}\cos{\Op{\phi}} 
+\sin^2{\beta}\sin^2{\Op{\theta}}\cos^2{\Op{\phi}}\Big)+\alpha_{\perp}\Big] \\
&&+ \frac{Q_Z\varepsilon^{3/2}(t)}{4} \left[3\left(\cos^2{\beta}\cos^2{\Op{\theta}} 
+ 2\cos{\beta}\sin{\beta}\cos{\Op{\theta}}\sin{\Op{\theta}}\cos{\Op{\phi}} 
+\sin^2{\beta}\sin^2{\Op{\theta}}\cos^2{\Op{\phi}}\right) + 1 \right].
\end{eqnarray}  
\end{widetext}

\subsection{Estimating $\kappa$ in the low-$\chi_D$ limit}\label{sec:Akappa}
We can use our model to predict at which value of the impact parameter, $b$, maximum excitation occurs. As a first step we take the derivative with respect to $\chi_D$ and setting it to zero. Since the maximum occurs at the same values of $\chi_D$ for $|c|^2$ as for $c$ itself we will consider the derivative of the absolute square with respect to $\chi_D$
\begin{widetext}
\begin{equation}
 \frac{\partial |c(\chi_D, \kappa)|^2}{\partial \chi_D} = 2\left(\frac{\pi}{2\sqrt{3}}\right)^2
 \chi_D\kappa^2\exp{\left(-2\kappa\sqrt{1 + \frac{\chi_D^2}{3}}\right)} 
 \left[1 - \frac{\chi_D^2}{3}\kappa\sqrt{1 + \frac{\chi_D^2}{3}}\right].
\end{equation}
\end{widetext}
Setting the partial derivative to zero we get
\begin{equation}
 \chi_D^2\kappa\sqrt{1 + \frac{\chi_D^2}{3}} = 3.
\end{equation}
As an estimate we can set the expression in the square root equal to one, from which we get the approximate relation
\begin{equation}\label{eq:chistar}
 \chi_{D*} \approx \sqrt{\frac{3}{\kappa}}
\end{equation}
at maximum population transfer. This expression allows us to estimate the value of $\chi_D$ at maximum population transfer, which can serve as an estimate if population transfer can be expected at the given scattering energy. Furthermore, we can use Eqs.~\eqref{eq:coulomb}, \eqref{eq:r0}, \eqref{eq:chiD} and \eqref{eq:chistar} to estimate the value of $b$ at which maximum population transfer occurs as a function of scattering energy for a given scattering pair, i.e. solve for $b$ at $\chi_D = \chi_{D*}$ The result is
\begin{equation}\label{eq:bstar2}
 b \approx \sqrt{\left(\sqrt{\frac{D}{B}\sqrt{\frac{\kappa}{3}}} - \frac{\change{e^2}}{2E}\right)^2 - \left(\frac{\change{e^2}}{2E}\right)^2}.
\end{equation}
\begin{figure}[tbp]
 \centering
 \includegraphics[scale=0.35]{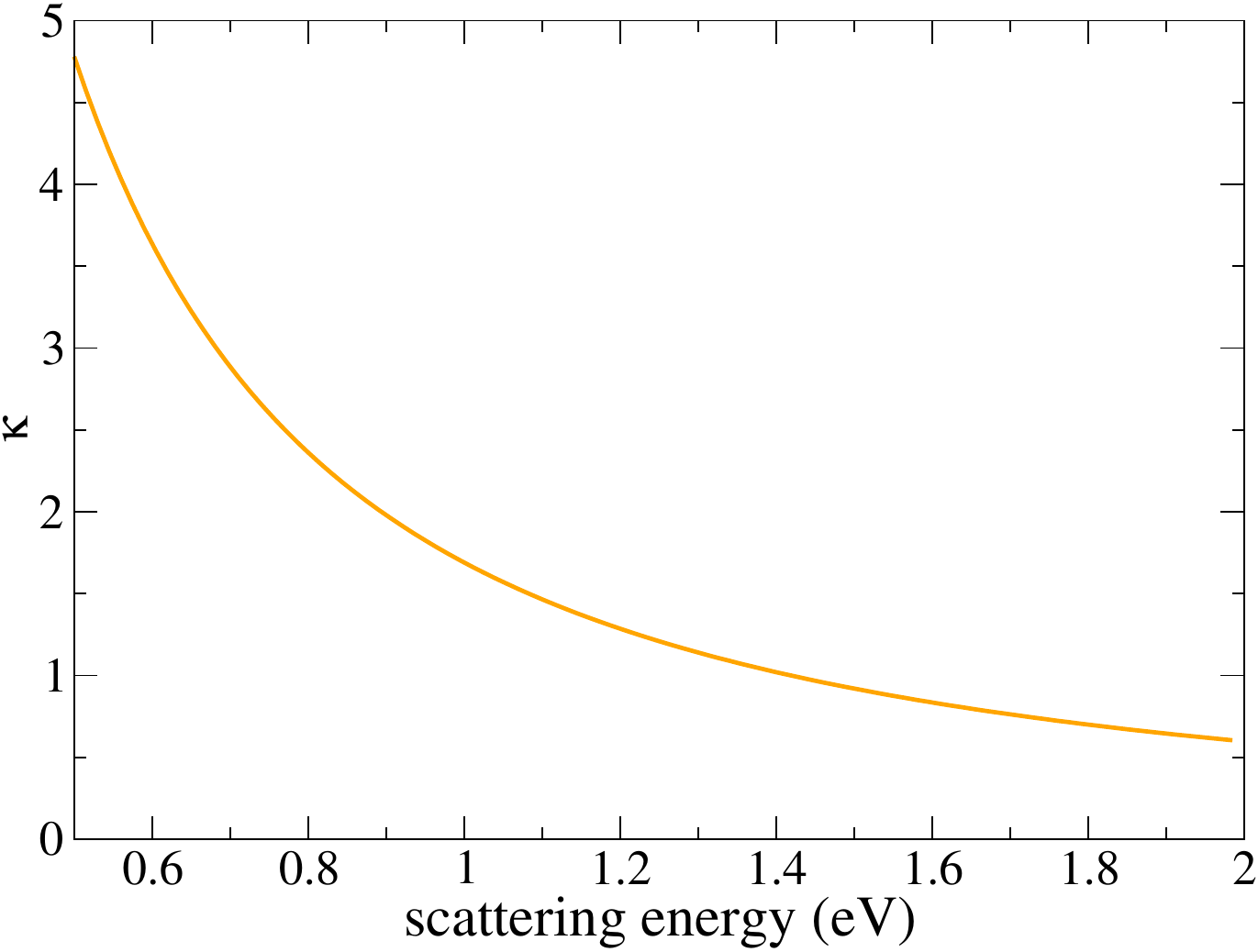}
 \caption{$\kappa$ at $b_*$, i.e. Eq.~\eqref{eq:chistar} as a function of the scattering energy.}
 \label{fig:kappaStar}
\end{figure}
It is of relevance to estimate when the adiabatic picture is applicable. Typically that is when the rotational time is short compared to the duration of the field, i.e. when $\kappa > 1$ (or even $ \gg 1$). This estimate can only be reliable when the internal rotational structure is not significantly altered by the field so as to leave $T_{rot} \propto B^{-1}$. This is the case for HD$^+$, as can be seen from Fig.~\ref{fig:EJ_Bnorm}. For HD$^+$ we therefore expect the adiabatic picture to be relevant only for $\kappa > 1$. It is seen from Fig.~\ref{fig:kappaStar} that $\kappa = 1$ at around $E \approx 1.5$ eV for HD$^+$ and decreasing with energy. We therefore estimate that the adiabatic picture is only relevant for scattering energies below $1.5$ eV. In the figure we see that the adiabatic picture should lose its relevance at even smaller scattering energies for MgH$^+$, in contrast to what we see from numerical calculations. However, unlike HD$^+$ the internal rotational states are significantly affected by the external field due to the coolant, and the simple estimate based on $\kappa > 1$ for adiabaticity cannot be convincingly applied. 
\medskip
\bibliography{refs}
\end{document}